\documentclass[longauth]{aa} % for the long lists of affiliations 
\usepackage{graphicx}
\usepackage{txfonts}
\usepackage{natbib}
\usepackage{multirow}
\usepackage{lscape}
\usepackage{longtable}
\def\chandra{{\em Chandra}}

\def\cm2{{cm$^{-2}$}}

\def\Hb{{H$\beta$}}
\def\Ha{{H$\alpha$}}
\def\oii{[\ion{O}{ii}]}
\def\oiii{[\ion{O}{iii}]}
\def\nii{[\ion{N}{ii}]}
\def\Hanii{{H$\alpha$}/[\ion{N}{ii}]}
\def\niiHa{[\ion{N}{ii}]/{H$\alpha$}}
\begin{document}
   \title{GLACE survey: OSIRIS/GTC Tuneable Filter H$\alpha$ imaging of the rich galaxy cluster ZwCl 0024.0+1652 at z\,=\,0.395}

   \subtitle{Part I -- Survey presentation, TF data reduction techniques and catalogue}
   
   \titlerunning{GLACE: OSIRIS/GTC TF \Ha\ imaging of ZwCl 0024.0+1652. Part I}

   \author{M. S\'anchez-Portal
          \inst{1,2}
          \and
          I. Pintos-Castro          
          \inst{3,4,8,2}
          \and
          R. P\'erez-Mart\'{\i}nez
          \inst{1,2}
          \and
          J. Cepa,
          \inst{4,3}
          \and
          A. M. P\'erez Garc\'{\i}a
          \inst{3,4}
          \and
          H. Dom\'{\i}nguez-S\'anchez
          \inst{29}
          \and
         A. Bongiovanni
          \inst{3,4}
          \and
         A. L. Serra
          \inst{23}
          \and
          E. Alfaro
          \inst{5}
          \and
          B. Altieri
          \inst{1}
          \and
          A. Arag\'on-Salamanca 
          \inst{6}
          \and
          C. Balkowski
          \inst{7}
          \and
          %M. Balogh
          %\inst{8}
          %\and
          A. Biviano
          \inst{9} 
          \and
          M. Bremer
          \inst{10}
          \and
          F. Castander
          \inst{11} 
          \and
          H. Casta\~neda
          \inst{12}
          \and
          N. Castro-Rodr\'{\i}guez
          \inst{3,4}
          \and
          A. L. Chies-Santos
          \inst{25}
          \and
          D. Coia
          \inst{1}
          \and
          A. Diaferio
          \inst{23,24}
          \and
          P.A. Duc
          \inst{13}
          \and
          A. Ederoclite
          \inst{21}
          \and
          J. Geach
          \inst{14}
          \and
          I. Gonz\'alez-Serrano
          \inst{15}
          \and
          C. P. Haines
          \inst{16}
          \and
          B. McBreen
          \inst{17}
          \and
          L. Metcalfe 
          \inst{1}
          \and
          I. Oteo
          \inst{26,27}
          \and
          I. P\'erez-Fourn\'on
          \inst{4,3}
          \and
          B. Poggianti
          \inst{18} 
          \and
          J. Polednikova
          \inst{3,4}
          \and
          M. Ram\'on-P\'erez
          \inst{3,4}
          \and
          J. M. Rodr\'{\i}guez-Espinosa
          \inst{4,3}
          \and
          J. S. Santos
          \inst{28}
          \and
          I. Smail
          \inst{19}
          \and
          G. P. Smith
          \inst{17}
          \and
          S. Temporin
          \inst{20}
          \and 
          I. Valtchanov
          \inst{1}
          }
   \authorrunning{M. S\'anchez-Portal et al.}
   \institute{European Space Astronomy Centre (ESAC)/ESA, P.O. Box 78, 28690 Villanueva de la Canada, Madrid, Spain\\
              \email{miguel.sanchez@sciops.esa.int}
              \and
              ISDEFE, Beatriz de Bobadilla 3, 28040 Madrid, Spain
              \and
              Instituto de Astrof\'{\i}sica de Canarias, E38205, La Laguna, Tenerife, Spain
              \and
              Universidad de La Laguna, Tenerife, Spain
              \and
              Instituto de Astrof\'{\i}sica de Andaluc\'{\i}a, CSIC, Granada, Spain
              \and
              School of Physics and Astronomy, University of Nottingham, U.K.
              \and
              GEPI, Observatoire de Paris \& CNRS,  Meudon, France
              \and
              Centro de Astrobiolog\'{\i}a, INTA-CSIC, Madrid, Spain
              %\and
              %Department of Physics and Astronomy, University of Waterloo, Canada
              \and
              INAF, Osservatorio Astronomico di Trieste, Italy
              \and
              Department of Physics, University of Bristol, U.K.
              \and
              Institut de Ci\`encies de l'Espai (CSIC), Barcelona, Spain
              \and
              Instituto Polict\'ecnico Nacional, M\'exico D.F., M\'exico
              \and
              Laboratoire AIM Saclay, CEA/IRFU, CNRS/INSU, Universit\'e Paris Diderot, France
              \and
              Department of Physics, McGill University, Montreal, Quebec, Canada
              \and
              Instituto de F\'{\i}sica de Cantabria, CSIC-Univ. de Cantabria, Santander, Spain
              \and
              Departamento de Astronom\'{\i}a, Universidad de Chile, Casilla 36-D, Correo Central, Santiago, Chile
              \and
              University College, Belfield, Dublin, Ireland
              \and
              INAF, Osservatorio Astronomico di Padova, Italy
              \and
              Institute for Computational Cosmology, Durham University, Durham DH1 3LE, U.K.
              \and
              Institute of Astro- and Particle Physics, University of Innsbruck, Austria
              \and
              Centro de Estudios de F\'{\i}sica del Cosmos de Arag\'on, Teruel, Spain  
              \and
              Dipartimento di Fisica, Universit\`a degli Studi di Milano, Milano, Italy
              \and
              Dipartimento di Fisica, Universit\`a di Torino, Torino, Italy
              \and
              Istituto Nazionale di Fisica Nucleare (INFN), sezione di Torino, Torino, Italy 
              \and
              Departamento de Astronomia, Instituto de Astronomia, Geof\'isica e Ci\^encias Atmosf\'ericas da USP, S\~ao Paulo, Brazil
              \and
              Institute for Astronomy, University of Edinburgh, Royal Observatory, Blackford Hill, Edinburgh
              \and
              European Southern Observatory, Garching, Germany
              \and
              INAF-Osservatorio Astrofisico di Arcetri, Florence, Italy
              \and
              Departamento de Astrof\'{\i}sica, Facultad de CC. F\'{\i}sicas, Universidad Complutense de Madrid, 28040, Madrid, Spain
             }

   \date{Received; accepted }

% \abstract{}{}{}{}{} 
% 5 {} token are mandatory
 
  \abstract
   {The cores of clusters at 0\,$\lesssim$\,z\,$\lesssim$\,1 are dominated by quiescent early-type  galaxies,
whereas the field is dominated by star-forming late-type galaxies. Clusters
grow through the accretion of galaxies/groups from the surrounding
field, which implies that galaxy properties, notably the star formation ability,
 are altered as they fall
into overdense regions. 
%Where do these changes occur and why? 
%Current
%observations in the local Universe, and hints from higher-z, suggest
%that the truncation of star formation occurs in groups far out from the
%cluster core before they accrete onto the cluster core. 
The critical
issues to understand this evolution are how the truncation of star
formation is connected to the morphological transformation and what
physical mechanism is responsible for these changes.  The GaLAxy Cluster Evolution Survey (GLACE)
%This study exploits the
%extensive information already available for these clusters (redshifts, multi-colour 
%broadband imaging, X-ray, mid-IR etc.)
%redshift information available for these clusters, as well as
%a wide-field imaging with HST ACS+WFPC2, multi-colour Subaru
%photo-redshifts, panoramic MIPS 24-um surveys and deep wide-field X-ray
%data. 
is conducting a thorough study on the variation of galaxy properties (star
formation, AGN activity and morphology) as a function of environment
%(galaxy, gas or dark matter) 
in a representative and %uniquely
well-studied sample of clusters. 
%This study will address the key
%questions about the physical processes acting upon the infalling
%galaxies during the course of hierarchical growth of clusters.
}
   {To address these
questions,  the GLACE survey is performing a deep panoramic survey of emission line galaxies (ELG), mapping a set of  
optical lines  (\oii, \oiii, \Hb\  and \Hanii\  when possible) in several
galaxy clusters at z\,$\sim$\,0.40, 0.63 and 0.86. 
%This study is addressing key
%questions about the physical processes acting upon the infalling
%emission-line galaxies (ELG) during the course of hierarchical growth of clusters. 
}
   { Using the Tunable Filters (TF) of the OSIRIS instrument
at the 10.4m GTC telescope, the GLACE survey applies the technique of TF tomography: for each line, a set of images are taken through the OSIRIS TF, each image tuned at a different wavelength (equally spaced), to cover a rest frame velocity range of several thousands km/s centred at the mean cluster redshift is scanned for the full TF field of view of 8\,arcmin diameter.  
}
   {
   Here we present the first results of the GLACE project, targeting the \Hanii\ lines in the intermediate redshift cluster ZwCl 0024.0+1652 at z\,=\,0.395. Two pointings, covering $\sim$\,2\,$\times$\,r$_{vir}$ 
%(the first centered at the cluster core and a second one displaced toward the North-West) 
have been performed. We discuss the specific techniques devised to process the TF tomography observations in order to generate the catalogue of cluster \Ha\ emitters, that contains more than 200 sources down to a star formation rate (SFR)\,$\lesssim$\,1\,M$_{\odot}$/yr. An ancillary broadband catalogue is constructed,  allowing us to discriminate line interlopers by means of colour diagnostics. The final catalog contains 174 unique cluster sources.
%By means of spectral energy distribution (SED) fitting to the broadband data, we derive the stellar masses of galaxies used to compute the specific SFR (sSFR) and also to correct the line fluxes for the underlying stellar absorption component. 
The AGN population is discriminated using different diagnostics and found to be $\sim$\,37\% of the ELG population. The median SFR of the star-forming population is 1.4\,M$_{\odot}$/yr. We have studied the spatial distribution of ELG, confirming the existence of two components in the redshift space. Finally, we have exploited the outstanding spectral resolution of the TF, attempting to estimate the cluster mass from ELG dynamics, finding M$_{200}$\,$=$\,(4.1\,$\pm$\,0.2)\,$\times$\,10$^{14}$\,M$_{\odot}\, h^{-1}$, in agreement with previous weak-lensing estimates. 
}
 {}

   \keywords{galaxies: clusters: individual: ZwCl 0024.0+1652 --
                galaxies: photometry --
                galaxies: star formation --
                galaxies: active
               }

   \maketitle
%
%________________________________________________________________

\section{Introduction}

\label{sec:intro}

It is well known that, while the cores of nearby clusters are dominated by red
early-type galaxies, a significant increase in the fraction of blue cluster galaxies is observed at z\,$>$\,0.2 \citep[the so-called Butcher-Oemler -BO- effect; ][]{Butcher1984}.  
An equivalent increase in obscured star formation (SF) activity has also been seen in mid- and far- IR surveys of
distant clusters \citep{Coia2005,Geach2006,Haines2009,Altieri2010}  as well as a growing population of AGN \citep[e.g. ][]{Martini2009,Martini2013}. In general, a strong evolution (i.e. increase of the global cluster star formation rate --SFR) is observed 
\citep[e.g. ][]{Webb2013,Koyama2011}.
Even focusing on a single epoch, aspects of this same
evolutionary trend have been discovered in the outer parts of clusters
where significant
changes in galaxy properties can be clearly identified such as
gradients in typical colour or spectral properties with clustercentric
distance \citep{Balogh1999,Pimbblet2001}  and in the
morphology-density relation \citep{Dressler1980,Dressler1997}.
In a hierarchical model of structure formation, galaxies assemble into
larger systems, namely galaxy clusters, as time progresses. 
It is quite likely that this
accretion process is responsible for a transformation of the properties
of cluster galaxies both as a function of redshift and as a function of
environment \citep{Balogh2000,Kodama2001}.

\cite{Haines2013} measured the mid-IR BO effect over the redshift range 0.0--0.4, finding a rapid evolution in the fraction of cluster massive ($M_K < -23.1$) luminous IR galaxies within r$_{200}$ and SFR\,$>$\,3\,M$_{\odot}$/yr that can be
modeled as $f_{SF}$\,$\propto$\,(1+z)$^n$, with n\,=\,7.6\,$\pm$\,1.1. The authors investigate the origin of the BO effect, finding that can be explained as a combination of a $\sim$\,3$\times$ decline in the mean specific-SFR of  
star-forming cluster galaxies since z\,$\sim$\,0.3 with a $\sim$\,1.5$\times$ decrease in number density. Two-thirds of this reduction in the specific-SFRs of star-forming cluster galaxies is due to the steady cosmic decline in the specific-SFRs among those field galaxies accreted into the clusters. The remaining one-third reflects an accelerated decline in the SF activity of galaxies within clusters. The slow quenching of SF in cluster galaxies is consistent with a gradual shut down of SF in infalling spiral galaxies as they interact with the cluster medium. 

%\cite{Haines2009} found a steady increase of the fraction of cluster massive luminous IR galaxies  that can be
%modeled as $f_{SF}$\,$\propto$\,(1+z)$^n$, with n\,=\,5.7$^{+2.1}_{-1.8}$ in the range 0.02\,$\leq$\,z\,$\leq$0.40. This evolution
%is stronger than that of $L^*_{IR}$ in both cluster and field; the excess is statistically associated with galaxies found at large
%cluster-centric radii, suggesting that the mid-IR BO effect can be explained by a combination of: {\em (i)} a  global decline %of the 
%SFR since z\,$\sim$\,1 and {\em (ii)} enhanced star formation in the infall regions of clusters and subsequent quenching after passage through
%the cluster. Nevertheless, the observed $f_{SF}$ lies systematically above the prediction derived from this scenario, suggesting that
%other processes (e.g. SF triggered by environmental processes or a gradual quenching) could play a role.  
Possible physical processes that 
have been proposed to trigger or inhibit the SF include \citep[e.g.][]{Treu2003}: {\it (i)} Galaxy-ICM interactions: ram-pressure stripping, thermal evaporation of the ISM, 
turbulent and viscous stripping, pressure-triggered SF. When a slow decrease of the SF is produced, these mechanisms are collectively labelled as 
starvation. {\it (ii)} Galaxy-cluster gravitational potential interactions:  tidal compression, tidal truncation. {\it (iii)} Galaxy-galaxy interactions: mergers 
(low-speed interactions), harassment (high-speed interactions).  Nevertheless,
it is still largely unknown whether the correlations of star-formation histories 
and large-scale structure are due to the advanced evolution 
in overdense regions, or to a direct physical effect on the star formation
capability of galaxies in dense environments \citep[e.g.][]{Popesso2007}.
This distinction can be made reliably if one has an accurate
measurement of star formation rate (or history), for galaxies spanning
a range of stellar mass and redshift, in different environments. 

The physical processes proposed above act on the cluster population of emission line galaxies (ELG; comprising both SF and AGN population);  
narrow-band imaging surveys are very efficient  to identify {\it all} the ELGs in a cluster. For instance, \cite{Koyama2010}  performed a MOIRCS narrow-band
\Ha\ and \textit{Akari} mid-IR survey of the cluster RX J1716.4+6708 at z\,=\,0.81, showing that both \Ha\ and mid-IR emitters avoid the
central cluster regions. A population of red SF galaxies (comprising both \Ha\ and mid-IR emitters) is found in medium-density environments
like outskirts, groups and filaments, suggesting that dusty SF  is triggered in the infall regions of clusters, implying a probable link 
between galaxy transition and dusty SF. The authors found that the mass-normalised cluster SFR declines rapidly since z\,$\sim$\,1 as $\propto$\,(1+z)$^6$ \citep[i.e. consistent with 
the results from ][ outlined above]{Haines2013}.  
``Classical'' narrow-band imaging surveys have therefore demonstrated to be a powerful tool, but suffer
from ambiguity about the true fluxes of detected sources and do not provide neither accurate membership nor dynamical
information about the population.  Attempting to overcome these limitations, but taking advantage of the power of narrow-band imaging, the  GaLAxy Cluster Evolution Survey (GLACE) has been designed as an innovative survey of ELGs and AGNs in a well-studied and well-defined sample of clusters, exploiting  the novel capabilities of tunable filters (TF) of the OSIRIS instrument \citep[Optical System for Imaging and low-Resolution Integrated Spectroscopy; ][]{Cepa2003,Cepa2005} at the 10.4m GTC telescope, to map a set of important optical emission lines by means of the technique of ``TF tomography'', scanning a range of the spectrum, i.e. a range of radial velocities around the cluster nominal redshift, at a fixed wavelength step.  

The main purpose of this paper is to present the GLACE project, along with some specific techniques devised to process TF data and the results of several simulations performed to assess the quality and performance of the observations and procedures developed. In addition, results from the \Hanii\ survey performed in the galaxy cluster ZwCl 0024+1652 (hereafter referred as Cl0024) at z\,$=$\,0.395 are presented. In sect. \ref{sect:glace}, the GLACE survey is outlined, including its main objectives, technical implementation and intended targets. Then, sect. \ref{sect:obs_data_reduction}
reports on the observations performed towards Cl0024, along with the procedures developed to reduce TF data within the GLACE survey. The method for flux calibration is also addressed, as well as Cl0024 ancillary broadband data and spectroscopic redshifts. In sect. \ref{sect:catalogue} the methods to derive the catalogue of ELG, including line wavelength estimation and rejection of contaminants (line emitters at different redshifts) are described. sect. \ref{sect:ha_flux} addresses the derivation of the \Ha\ and \nii\ fluxes and possible explanations to the absorption-like features observed in some cases in the spectral data. In sect. \ref{sect:redshift} we discuss the spatial and redshift distribution of the cluster galaxies. sect. \ref{sect:Ha_LF} is dedicated to the \Ha\ luminosity function of Cl0024 and its comparison with previous work \citep{Kodama2004}. In sect. \ref{sect:SF_AGN}, the discrimination between star-forming (SF) galaxies and AGNs is discussed. Finally, sect. \ref{sect:cluster_dynamics} investigates the possibility of studying the dynamical properties of clusters of galaxies by means of emission (rather than absorption) lines.

This is the first of a series of papers on the GLACE targets. A thorough discussion on the SF properties of Cl0024 derived from the \Ha\ line and their comparison with mid- and far-IR results,  as well as discussion on the results derived from the rest of lines targeted will be addressed in forthcoming papers (P\'erez-Mart\'inez et al., in preparation). The infrared-derived SF properties of the young galaxy cluster RX J1257.2+4738 at z = 0.866  (see GLACE target list below) have been discussed in \cite{Pintos-castro13}. The GLACE \oii\ OSIRIS TF survey of this cluster and the morphological properties of the SF and AGN population are being addressed in Pintos-Castro et al. (in preparation). 

Unless otherwise specified, throughout this paper we assume a Universe with H$_0$\,=\,70\,km\,s$^{-1}$\,Mpc$^{-1}$ , $\Omega_{\Lambda}$\,=\,0.7 and $\Omega_{m}$\,=\,0.3. 

%__________________________________________________________________

\section{The GLACE Project}
\label{sect:glace}

\begin{figure}
   \centering
   \includegraphics[width=9cm]{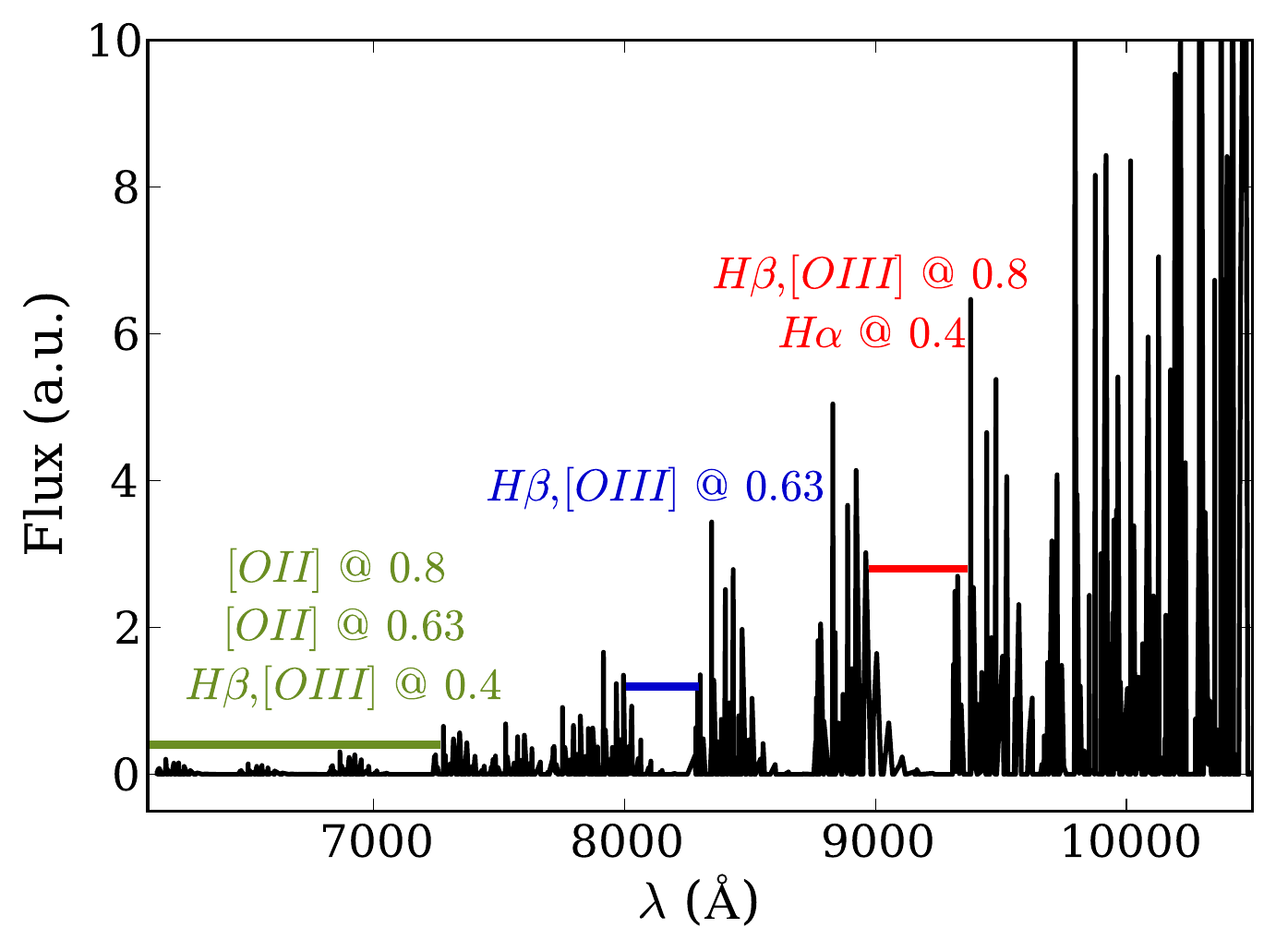}
   \vspace{-0.5cm}
      \caption{Night sky spectrum from \cite{Rousselot2000} and GLACE windows within the OH bands. The locii of the strongest rest-frame emission lines at different redshifts are shown next to the corresponding spectral windows. The observations within this paper have been performed within the \Ha\ window at z\,=\,0.4.
              }
         \label{glace_windows}
   \end{figure}

The GLACE programme (PIs. Miguel S\'anchez-Portal \& Jordi Cepa) is undertaking a
panoramic census of the star formation and AGN activity within a sample of
clusters at three redshift bins defined by windows relatively free of strong atmospheric OH emission lines (Fig. \ref{glace_windows}):  z\,$\sim$0.40, 0.63 and 0.86, mapping the strongest rest-frame optical emission lines:
H$\alpha$ (only at z\,$\sim$\,0.4), H$\beta$, [O{\sc ii}]3727 and [O{\sc iii}]5007; the sections below describe the main goals of the GLACE project and the technical implementation of the project.

\subsection{GLACE goals}
\label{subsect:goals}

The GLACE project is aimed at comparing the maps of ELGs with the
structures of the targeted clusters (as traced
by galaxies, gas and dark matter) to address 
several crucial issues:

\noindent (1) Star formation in clusters: We will determine how the star formation properties of galaxies relate to
their position in the large scale structure. This will provide a key
diagnostic to test between different models for the environmental
influence on galaxy evolution. Each mechanism 
%(see Treu et al.\ 2003)
is most effective in a different environment \citep[generally depending on the cluster-centric distance, e.g.][]{Treu2003}, leaving a footprint in
the data.  We are mapping the extinction-corrected star formation through H$\alpha$ and [O{\sc ii}], over a
large and representative region in a statistically useful sample of
clusters. The survey has been designed to reach SFR $\sim$2 M$_\odot$/yr (i.e. below
that of the Milky Way) with 1 magnitude of extinction at H$\alpha$\footnote{f$_{H\alpha}$\,=\,1.89\,$\times$\,10$^{-16}$\,erg\,s$^{-1}$ at z\,=\,0.4 using standard 
SFR$-$luminosity conversion factors \citep{Kennicutt1998}}.  Our first results
(see sect. \ref{sect:catalogue}) show that we are achieving this goal.  Another important question that can be addressed is related with the kind of
galaxies (type/mass) forming stars within clusters.  
%It has been suggested \cite{Couch01}
%a morphological transformation from infalling spirals to
%cluster S0s to explain a rapid decrease in S0 fraction with redshift in
%cluster cores.  
The study of the morphology of confirmed ELGs allows investigating
 the connection between the truncation
of star formation and the morphological transformation of infalling
galaxies \citep[e.g.][]{Poggianti1999}. In addition, by determining the total
integrated star formation rates, we can construct a
star-formation history for galaxies in clusters, as has
been done in the field \citep[e.g.][]{Madau1998}.

\noindent (2) The role of AGNs:  
Whether the fraction of AGNs is
environment-dependent or not is a matter of debate: while some results \cite{Miller2003} point towards 
a lack of dependency of the AGN fraction with the local galaxy density, 
other authors  \citep[e.g.][]{Kauffmann2004} conclude that high-luminosity AGNs do avoid high-density
regions. Such a lack of AGNs in clusters, as compared to the field, may
be related to the evolution of galaxies as they enter the cluster
environment.  On the other hand, claims for an enhanced
fraction of AGNs in groups \citep[e.g.][]{Popesso2006} point at AGN-stimulating
processes such as galaxy--galaxy interactions or mergers that are
particularly effective in the low-velocity dispersion and (relatively)
high-density environments typical of groups (and filaments).
 
Regarding redshift evolution, \cite{Martini2009,Martini2013} have studied the fraction of luminous AGNs in clusters up to z\,$\simeq$\,1.5, finding that it is 30 times greater at 1\,$<$\,z\,$<$\,1.5 than the fraction found in clusters at z\,$\sim$\,0.25 and more than one order of magnitude larger than  that found at z\,$\sim$\,0.75, indicating that the AGN population in clusters has evolved more rapidly than the field one, in analogy with the faster evolution of cluster SF galaxies with respect to field.

A complete census of the AGN population in clusters at different
redshifts, and in cluster regions characterized by different mean
densities and velocity dispersions, can help us to constrain the physics
behind the onset of the AGN activity in galaxies. Our survey is 
sensitive to the AGN population within clusters, discriminated
from pure star formation by means of standard diagnostics, e.g. BPT \citep{Baldwin1981} and EW$\alpha$n2 \citep{CidFernandes2010} diagrammes.

\noindent (3)  The study of the distribution of galaxy metallicities 
with cluster radii is another powerful mean to investigate 
evolution within clusters. As galaxies fall into clusters and travel toward
the cluster center along their radial orbits, they interact with 
the intra-cluster medium (ICM) and other cluster galaxies, thus getting progressively 
stripped of their gas reservoir. This process is expected to influence their
metal abundances and possibly generate a metallicity gradient across a cluster.
Not much is yet known on the metallicities of emission-line galaxies in
clusters and how it varies with cluster-centric distances. GLACE will
allow to derive extinction-corrected metallicities using N2 \citep{Denicolo2002}, R23 
\citep{Pagel1979}, and O3N2 \citep{Alloin1979}, where N2 and O3N2
will be used to break the R23 degeneracy and to assess possible 
differences between N and O abundances vs metallicity. 

The survey addresses many other interesting topics: for instance the cluster accretion history can be traced
by studying the census of ELGs at  different cluster-centric distances.  
In addition,  the survey will provide an accurate assessment of cluster membership, without 
the need of a spectroscopic follow-up. 

%%
%%______________________________________________________________

\subsection{GLACE implementation and TF principles}
\label{subsect:implementation}

Regarding the technical implementation, the GLACE survey exploits the outstanding characteristics of the tunable filters \citep[TF; ][]{Gonzalez2014,Cepa2005,Cepa2003}; these are special Fabry-Perot interferometers in which the two plane parallel transparent plates (covered with high-reflectivity, low-absorption coatings) are operated at much smaller spacing than the traditional ones (typically few microns). Hence, a broadened central interference region is produced (known as the Jacquinot spot, defined as the region over which the change in wavelength does not exceed by $\sqrt{2}$ times the FWHM). Moreover, controlled by means of a stack of piezo-electric transducers, the  plate spacing can be varied with high accuracy over a large range. Therefore, a wide spectral region is accessible with a moderate spectral resolution (of the order of 10--20\,\AA\ FWHM): in the case of the OSIRIS red TF, the spectral range is 6510--9345\,\AA. 

The TF transmission profile (Airy function) is periodic with the incident light wavelength. For a beam entering the TF at incidence angle $\theta$, the condition for a maximum transmission (constructive interference) is:
\begin{equation}
m\lambda = 2\mu d \cos \theta
\label{interference_eq}
\end{equation}

\noindent where \textit{m} is the integer order of interference, $\mu$ is the refractive index of the medium in the cavity between the plates (usually air, $\mu$\,=\,1) and \textit{d} is the plate separation (gap). In order to select just one wavelength and order, additional intermediate-band filters, known as order sorters, are required. Since light coming from targets at increasing distance from the OSIRIS optical centre reaches the TF at increasing incidence angle, there is a progressive shift towards the blue as the distance \textit{r} of the source to the optical centre increases. For the OSIRIS red TF, the dependency of the transmitted wavelength on the radial distance is given by \cite{Gonzalez2014}:

\begin{equation}
\lambda = \lambda_0 - 5.04 r^2 + a_3(\lambda) r^3
\label{radial_dep}
\end{equation}

\noindent where $\lambda_0$ is the central wavelength tune in \AA, \textit{r} is the distance to the optical centre in arcmin and $a_3(\lambda)$ is an additional term expressing the wavelength dependency of the coatings, given by:

\begin{equation}
a_3(\lambda) = 6.0396 - 1.5698 \times 10^{-3} \lambda + 1.0024 \times 10^{-7} \lambda^2
\label{a_3}
\end{equation}

Within a wavelength period, the TF transmission profile $T(\lambda)$ can be approximated by the expression:

\begin{equation}
T\left(\lambda\right) \simeq \left(1 + \left(\frac{2 \left(\lambda -\lambda_0\right)}{\Delta_{FWHM}}\right)^2\right)^{-1}
\label{eq:transmission}
\end{equation}

\noindent where $\lambda_0$ is the wavelength at which the TF is tuned and $\Delta_{FWHM}$ is the TF  FWHM bandwidth.

Within the GLACE survey, we have applied the technique of TF tomography \citep{Jones2001, Cepa2013}:
for each line, a set of images are taken 
through the OSIRIS TF, each image tuned at a different wavelength (equally spaced), 
so that a rest frame velocity range of several thousands km/s (6500 km/s for our first target) centred at the mean 
cluster redshift is scanned for the full TF field of view of 8\,arcmin in diameter. 
Additional images are taken to compensate for the blueshift of the wavelength from centre 
to the edge of the field of view (as given in Eq. \ref{radial_dep}).
Finally, for each pointing and wavelength tuned, 
three dithered exposures allow correcting for etalon diametric 
ghosts, using combining sigma clipping algorithms. 

The TF FWHM and sampling (i.e.: the wavelength interval between 
consecutive exposures)
%; see Figure \ref{Fig:TFpluslines}) 
at H$\alpha$ are of 12 and 6\,\AA, 
respectively, to allow deblending H$\alpha$ from [N{\sc ii}]$\lambda$6584.
with an accuracy better than 20$\%$ \citep{Lara2010}.  
For the rest of the lines, the largest available TF FWHM, 20\,\AA\ is applied, 
with sampling interval of 10\,\AA. These parameters also allow 
a photometric accuracy better than 20$\%$ according to simulations performed within the OSIRIS team.
The same pointing
positions are observed at every emission line. In order to trace the relation between SF and environment in a wide range of local densities, We have required to cover $\simeq$2 Virial radii (some 4 Mpc) within the targeted clusters. This determines the
number of OSIRIS pointings (two pointings at 0.40 and 0.63 and just one at 0.86).

%The original GLACE sample submitted to ESO in period 82B  included nine clusters (three at each redshift bin). Eventually one of the clusters in the low (0.4) bin, Zw Cl 0024+1654 

The intended GLACE sample includes nine clusters, three in each redshift bin. Currently we have been awarded with observing time and completed the observations of clusters: ZwCl 0024.0+1652 and RX J1257.2+4738. Tab. \ref{glace_sample} outlines the sample and the current status of the observations.

%%XXXXXXXX here goes the table with the sample!!
\begin{table*}
\caption{GLACE sample and status of the observations}             % title of Table
\label{glace_sample}      % is used to refer this table in the text
\centering                          % used for centering table
\begin{tabular}{lcccp{6cm}}        % centered columns (4 columns)
\hline\hline                 % inserts double horizontal lines
Name &RA(J2000)&Dec(J2000)&z&Status\\    % table heading 
\hline                        % inserts single horizontal line
ZwCl 0024.0+1652&00 26 35.7&+17 09 45& 0.395& Completed; programmes GTC63-09B, GTC8-10AGOS, GTC47-10B \& GTC75-13B\\
Abell 851 &09 42 56.6 & +46 59 22 & 0.407 & Planned\\
RX J1416.4+4446 & 14 16 28.7 & +44 46 41 &0.40& Planned\\
XMMLSS-XLSSC 001& 02 24 57.1&-03 48 58&0.613&Planned\\
MACS J0744.8+3927&07 44 51.8& +39 27 33&0.68&Planned\\
Cl J1227.9-1138&12 27 58.9&-11 35 13&0.636&Planned\\
XLSSC03&02 27 38.2&-03 17 57.0&0.839&Planned\\
RX J1257.2+4738 & 12 57 12.2&+47 38 07&0.866&Completed; ESO/GTC programme 186.A-2012\\ 
Cl 1604+4304 & 16 04 23.7& +43 04 51.9 &0.89& Planned; Cl 1604 supercluster\\
\hline                                   %inserts single line
\end{tabular}
\end{table*}
%

%Extinctions can be estimated from the ratio of the Balmer lines H$\alpha$/H$\beta$.
%Photo-z obtained from broad band images will allow removing outliers. 
%These are expected to be mainly foreground emission line galaxies at 
%other redshifts. When observing H$\alpha$ at z = 0.40, 
%[O{\sc iii}]$\lambda$500.7nm emitters at z = 0.83 and [O{\sc ii}]$\lambda$372.7nm 
%emitters at z = 1.46 can be observed as well. However, at the 
%limiting fluxes considered here (see below) these contaminants will not be
%very significant and can easily be discriminated via photometric 
%redshifts. 

\section{Observations and data reduction}
\label{sect:obs_data_reduction}

Two OSIRIS/GTC pointings using the red TF were planned an executed towards Cl0024. The first one (carried out in GTC semesters 09B, 10A and 13B; hereafter referred as ``centre position'') targeted the \Hanii, \Hb\ and \oiii\ lines. The observations were planned to keep the cluster core well centered within CCD1\footnote{The OSIRIS detector mosaic is composed of two 2048\,$\times$\,4096 pixel CCDs abutted, with a plate scale of 0.125\,arcsec/pixel. See \cite{Cepa2005} and the OSIRIS Users' manual at \texttt{http://www.gtc.iac.es/instruments/osiris/media}}. The second pointing (hereafter referred as ``offset position'') was carried out in semesters 10B and 13B and targeted the same emission lines. This second pointing was offset by $\Delta\alpha$\,=\,-2.3\,arcmin, $\Delta\delta$\,=\,+2.5\,arcmin (i.e. some 3.4\,arcmin in the NW direction). A summary of the observations is presented in Tab.~\ref{observations}. Within the scope of this paper, we shall present the results from the \Hanii\ observations.  %while the discussion of the results derived from the rest of lines targeted will be addressed in a forthcoming paper (P\'erez-Mart\'inez et al., in preparation). 
The total on-source exposure time is 5.1 and 2.9 hours at the central and offset pointings, respectively.

%%XXXXXXXX observations
%\begin{table*}
%\caption{Summary of OSIRIS/GTC TF observations of the  Cl0024}             % title of Table
%\label{observations}      % is used to refer this table in the text
%\centering                          % used for centering table
%\begin{tabular}{lcccccccc}        % centered columns (4 columns)
%\hline\hline                 % inserts double horizontal lines
%\multirow{2}{*}{Position}&RA(J2000)&Dec(J2000)&\multirow{2}{*}{Line}&Spectral range&\multirow{2}{*}{Num. scans}&Average exposure & Average seeing\\
%        & hh:mm:ss.s&dd:mm:ss&  & (\AA)  & & time (sec) & FWHM (arcsec)\\    % table heading 
%\hline                        % inserts single horizontal line
%\hline                                   %inserts single line
%\end{tabular}
%\end{table*}

%  Cl0024 observations
    %\begin{landscape}
         \begin{table*}
         \caption{Log of the OSIRIS/TF observations of the region centred around the \Ha\ emission line in the Cl0024 cluster.} 
         \label{observations}    
         \begin{center}
         \begin{tabular}{c c r@{ }c@{ }p{5mm} c c c c}
         \hline\hline              
		%\multicolumn{9}{c}{\textbf{Cl0024+1654} $\quad$   \textbf{R.A.}\,$00^h 26^m 35^s$, \textbf{Dec.}\,$+17^{\circ}09'43''$} \\
		%\cmidrule(l{6mm}r{2mm}){2-8}
		
		\multicolumn{9}{c}{\textbf{Centre Position}} \\
		\hline 		%\toprule
         $\lambda_{0,i}$ & \multirow{2}{*}{OS Filter} & \multicolumn{3}{c}{\multirow{2}{*}{Date}} & Seeing & \multirow{2}{*}{N$º$ Steps} & \multirow{2}{*}{N$º$ Exp.} & Exp. Time \\
        (nm) &  &  &  &  & ($''$) &  &  & (s) \\ 
		\hline 		
		 904.74 & f893/50 & 2009 & Dec & 05 & 0.7\,--\,0.9 & 6 & 3 & 53 \\
         &  & 2010 & Aug & 17 & 0.8 & 6 & 3 & 85 \\
         908.34 & f902/40 & 2010 & Aug & 21 & 0.8 & 11 & 3 & 60 \\
         908.74 & f902/40 & 2009 & Dec & 05 & 0.7\,--\,0.9 & 2 & 3 & 53 \\
         909.54 & f902/40 & 2009 & Nov & 25 & 0.9\,--\,1.1 & 9 & 3 & 53 \\
         914.94 & f910/40 & 2009 & Nov & 25 & 0.9\,--\,1.1 & 9 & 3& 53 \\
         &  & 2010 & Aug & 01 & 0.8 & 14 & 3 & 60 \\
         920.34 & f910/40 & 2009 & Dec & 05 & 0.6\,--\,0.8 & 5 & 3 & 53 \\
         923.34 & f919/41 & 2009 & Dec & 05 & 0.6\,--\,0.8 & 4 & 3 & 53 \\
         & & 2010 & Aug & 18 & 0.8 & 3 & 3 & 60 \\
         925.14 & f919/41 & 2010 & Aug & 18 & 0.8 & 5 & 3 & 100 \\
         928.14 & f919/41 & 2010 & Aug & 18 & 0.8 & 3 & 3 & 120 \\
         929.94 & f919/41 & 2010 & Aug & 19 & 0.9 & 3 & 3 & 190 \\
         931.74 & f923/34 & 2010 & Aug & 19 & 0.9 & 2 & 3 & 170 \\
         932.94 & f923/34 & 2010 & Aug & 21 & 0.8 & 3 & 3 & 120 \\
		\hline %\bottomrule
        & & & & & & & & \\
		\multicolumn{9}{c}{\textbf{Offset Position}} \\
		\hline 		%\toprule
         $\lambda_{0,i}$ & \multirow{2}{*}{OS Filter} & \multicolumn{3}{c}{\multirow{2}{*}{Date}} & Seeing & \multirow{2}{*}{N$º$ Steps} & \multirow{2}{*}{N$º$ Exp.} & Exp. Time \\
        (nm) &  &  &  &  & ($''$) &  &  & (s) \\ 
        \hline
		 904.74 & f893/50 & 2010 & Oct & 01 & $<$1.0 & 6 & 3 & 60 \\
         908.34 & f902/40 & 2010 & Nov & 08 & 1.0\,--\,1.2 & 11 & 3 & 60 \\
         914.94 & f910/40 & 2010 & Sep & 20 & $<$0.8 & 14 & 3& 60 \\
         923.34 & f919/41 & 2010 & Nov & 08 & 0.9\,--\,1.2 & 3 & 3 & 60 \\
         925.14 & f919/41 & 2010 & Nov & 08 & 0.9\,--\,1.2 & 5 & 3 & 100 \\
         928.14 & f919/41 & 2010 & Nov & 08 & 0.9\,--\,1.2 & 3 & 3 & 120 \\
         929.94 & f919/41 & 2010 & Nov & 08 & 1.0\,--\,1.2 & 3 & 3 & 120 \\
         932.94 & f923/34 & 2010 & Oct & 01 & $<$1.0  & 2 & 3 & 120 \\
        
		\hline %\bottomrule
         \end{tabular}
         \end{center}
         \end{table*}

The \Hanii\ spectral range 9047--9341\,\AA\  was covered by 50 evenly spaced scan steps\footnote{three scan steps, at 9317.4 and 9323.4 and 9341.4\,\AA\ were accidentally omitted in the observations of the offset positions; hence only 47 slices are available for that pointing.} ($\Delta\lambda$\,=\,6\,\AA). Taking into account the radial wavelength shift described by Eq.~\ref{radial_dep}, the spectral range sampled over the entire field of view (of 8\,arcmin diameter) is somewhat smaller, 9047--9267\,\AA\ (the ranges 9267--9341\,\AA\ and 8968--9047\,\AA\ are partially covered in the central and external regions of the field of view, respectively).   
%with a total of 5.15 hours of on-source integration time. 
At each TF tune, three individual exposures with an ``L'' shaped dithering pattern of 10\,arcsec amplitude were taken (in order to allow the removal of fringes and to ease the identification of diametric ghosts; the amplitude was chosen similar to the gap between the detectors). While this observing strategy has revealed useful for removing the fringing patterns that are specially evident beyond $\lambda$\,$\simeq$\,9300\,\AA, it introduces an additional complexity since the position of a source within the CCD varies in each dither position, and therefore, the wavelength at which it is observed due to the radial wavelength shift experienced in TFs.  

\subsection{TF data reduction procedures}
\label{TFreduction}

 The data reduction was performed using a version of the {\sc tfred} package \cite{Jones2002} modified for OSIRIS by A. Bongiovanni (Ram\'on-P\'erez et al. in prep.) and private {\sc iraf}\footnote{{\sc iraf} is distributed by the National Optical Astronomy Observatory, which is operated by the Association of Universities for Research in Astronomy (AURA) under cooperative agreement with the National Science Foundation.} and {\sc idl} scripts written by our team. The basic reduction steps were carried out using standard {\sc iraf} procedures and included bias subtraction and flat-field normalization; the next reduction step was a TF-specific one, namely the removal of the diffuse, optical-axis centred sky rings produced by atmospheric OH emission lines as a consequence of the radial-dependent wavelength shift described by Eqs. \ref{radial_dep} and \ref{a_3}; to this end, the  {\sc tfred} task \texttt{tringSub2} was used. It corrects each individual exposure by means of a background map created by computing the median of several dithered copies of the object-masked image. Fringing was also removed when required using the dithered images taken with the same TF tune. Then, the frames were aligned and a deep image obtained by combining all individual exposures of every scan step. This combination was done by applying a median filter. While this could potentially lead to the loss of line emitters with very low-continuum level, it is the best method to remove spurious features as ghosts. When compared with the deep image obtained adding up all the individual scans, applying a simple minimum-maximum rejection filter, we have observed that, at most, one line emitter per CCD gets lost but more than one-hundred spurious features are effectively removed. The astrometry was performed in the resulting deep image, using {\sc iraf} standard tasks (\texttt{ccxymatch} and \texttt{ccmap}). The sky position of reference objects were gathered from the USNO B1.0 catalogue \citep{Monet2003} in the centre position, while for the offset one we obtained better results using the 2MASS catalogue \citep{Skrutskie2006}; the achieved precision was in both cases equal or better than 0.3\,arcsec r.m.s. (i.e. of the order of the binned pixel size). The  deep images were used to extract the sources by means of the SExtractor package \citep{Bertin1996}. The number of sources detected above $3\sigma$ are 931 and 925 in the centre and offset positions, respectively (after removing a number of clearly spurious sources appearing  at the edges of the detectors). Since there is a quite large overlap between both positions, we found 374 common sources after matching both source catalogues using {\sc topcat} \citep{Taylor2005}. These common sources were used as a test of the relative consistency of our astrometry: 245 sources (65\%) were found within a match radius of 0.5\,arcsec (i.e. consistent with the quoted accuracy), 88 (24\%) within 0.75\,arcsec and 15 (4\%) within 1.0\,arsec radius. A number  of sources (26 objects, i.e. 7\%) were matched at larger radii (around 1.5\,arcsec) but these sources were always found at the edges of the images, were the OSIRIS field suffers a larger distortion.

The catalogue of detections contains 1482 unique sources. For each detected source and scan step, the best possible combination of individual images, i.e. the best combinations of TF tune and dither position at the location of the source was computed. In practice, we deemed as ``best combination'' algorithm the selection of all the images for which the TF wavelength at the position of the source lies within a range of $\pm$3\,\AA\ (i.e. half scan step) of the given one. 
For each of these combinations, a synthetic  equivalent filter transmission profile was derived by adding up the transmission profiles of all the images entering the combination and fitting to the result the function given in Eq.~\ref{eq:transmission}. These synthetic profiles allowed us to verify that our combination approach does not introduce a significant error neither in the wavelength of the central position (less than 1\,\AA\ maximum) nor in the FWHM (the average equivalent FWHM is 12.7\,\AA\  with a deviation of 0.4\,\AA).
The output combined image is used to determine the flux at this specific scan step and source position by means of SExtractor. The resulting ``pseudo-spectra'' consist of 50 (47) tuples ($\lambda$ at source position, flux). The FWHM of the equivalents (synthetic) TF Airy transmission profiles derived at each source position and TF tune were also included in each pseudo-spectrum file. A pseudo-spectrum should not be confused with a standard spectrum produced by a dispersive system: the flux at each point of the pseudo-spectrum is that integrated within a filter passband centred at the wavelength of the point; therefore, mathematically a pseudo-spectrum is the  convolution of the source spectrum with the TF transmission profile.

\subsection{Flux calibration}

\label{subsect:flux_calib}

The flux calibration of each TF tune has been carried out in two steps: first, the total efficiency $\epsilon(\lambda)$  of the system (telescope, optics and detector) should be derived; it is computed as the ratio of the measured to published flux $F_m(\lambda)/F_p(\lambda)$ for a set of exposures of spectrophotometric standard stars (Tab.~\ref{standards}) taken in photometric conditions within a range of tunes compatible with that of the cluster observation (ideally at the same tunes). The fluxes of the standards are measured by aperture photometry, and the exact wavelengths at the positions of the star are derived from Eqs. \ref{radial_dep} and \ref{a_3}. The published fluxes are also derived at these wavelengths by means of a polynomial fit to the tabulated fluxes (see references in Tab.~\ref{standards}). Then, measured fluxes in engineering units (ADU) are converted to physical units (ergs\,s$^{-1}$\,cm$^{-2}$\,\AA\ $^{-1}$)  using the expression:

\begin{table}
\caption{Spectrophotometric standard stars}             % title of Table
\label{standards}      % is used to refer this table in the text
\centering                          % used for centering table
\begin{tabular}{lcp{2.4cm}c}        % centered columns (4 columns)
\hline\hline                 % inserts double horizontal lines
Name & m($\lambda$) & Reference & Position \\ 
\hline                        % inserts single horizontal line
G157--34 & 15.35(5400) & \cite{Filippenko1984} &  offset \\
G191--B2B & 11.9(5556) & \cite{Oke1990} & offset \\
Ross 640 & 13.8(5556) & \cite{Oke1974} & centre \\
\hline                                   %inserts single line
\end{tabular}
\end{table}

    \begin{equation}
    \label{eq:fm}
         F_m(\lambda)=\frac{g\,K(\lambda)\,E_{\gamma}(\lambda)}{t\,A_{tel}\,\delta\lambda_{e}}F_{ADU}(\lambda)
    \end{equation}
    
\noindent where $g$ is the CCD gain in e$^-$ADU$^{-1}$, $E_{\gamma}(\lambda)$ is the energy of a photon in ergs, $t$ is the exposure time in seconds, $A_{tel}$ is the area of the telescope primary mirror in cm$^2$, $\delta\lambda_{e}$ is the effective passband width\footnote{$\delta\lambda_{e} = \frac{\pi}{2} FWHM_{TF}$ } in \AA, and $K(\lambda)$ is the correction for atmospheric extinction, 
    \begin{equation}
         K(\lambda)=10^{0.4\,k(\lambda) \, \left\langle \chi \right\rangle}
    \end{equation}
\noindent dependent on the extinction coefficient $k(\lambda)$ and the mean airmass $\left\langle \chi \right\rangle$ of the observations. In our case, we estimated  $k(\lambda)$ by fitting the extinction curve of La Palma\footnote{\texttt{http://www.ing.iac.es/Astronomy/observing/\\manuals/ps/tech\_notes/tn031.pdf}}.
in the wavelength range of interest.

The uncertainty in the efficiency has been computed by error propagation, taking into account the errors in the measured fluxes (that in turn include terms to cope with the error of the aperture photometry and the uncertainty of the wavelength tune) and those of the published ones.

The efficiency $\epsilon(\lambda)$  (sampled with 9 tunes at position A and 19 tunes at position B) must be then fitted to an analytical function of $\lambda$ in order to perform the calibration at the wavelength of each tune and source. In both cases, the best solution has been a constant efficiency for $\lambda$\,$\leq$\,9270\,\AA\ and a linear decreasing dependency at longer wavelengths (Tab.~\ref{tab:Cl0024eff}).

\begin{table}
\caption{Efficiencies for Cl0024 observations}             % title of Table
\label{tab:Cl0024eff}      % is used to refer this table in the text
\centering                          % used for centering table
\begin{tabular}{lccc}        % centered columns (4 columns)
\hline\hline                 % inserts double horizontal lines
Position & $\lambda$\,$\leq$\,9270\,\AA & \multicolumn{2}{c}{$\lambda$\,$>$\,9270\,\AA}\\ 
 & $\left\langle \epsilon \right\rangle$ & Zero point & Slope \\ 
\hline                        % inserts single horizontal line
%Centre & 0.1779\,$\pm$\,0.0236 & 9.8166\,$\pm$\,2.3318 & -0.0010\,$\pm$\,0.0002\\ 
Centre & 0.1779\,$\pm$\,0.0236 & 10.2497\,$\pm$\,1.2181 & -0.0011\,$\pm$\,0.0001\\ 
Offset & 0.1993\,$\pm$\,0.0035 & 12.8566\,$\pm$\,0.9074 & -0.0014\,$\pm$\,0.0001\\
\hline                                   %inserts single line
\end{tabular}
\end{table}

The second step is to convert the measured flux in ADU of each source at each tune $i$ to physical units  (ergs\,s$^{-1}$\,cm$^{-2}$\,\AA\ $^{-1}$) by means of the expression:

    \begin{equation}
    \label{eq:flambda}
        f(\lambda)_i = \frac{g\,K(\lambda)\,E_{\gamma}(\lambda)}{t\,A_{tel}\,\delta\lambda_{e}\,\epsilon(\lambda)}f_{ADU,i}
    \end{equation} 

\noindent where  $\epsilon(\lambda)$ is the total efficiency computed above and the remaining terms are as in Eq. \ref{eq:fm}. The flux errors are computed again by propagation, taking into account the efficiency errors derived above and the source flux measurement uncertainty  computed by the {\sc tfred} \texttt{tspect} task as:

\begin{equation}
\label{eq:flux_err_tspect}
\Delta f = \sqrt{A_{pix}\sigma^2 + f / g} 
\end{equation}

\noindent where $A_{pix}$ is the measurement aperture area in pixels, $\sigma$ is the standard deviation of the background noise and $g$ is the gain in in e$^-$ADU$^{-1}$.

\subsection{Ancillary data}

Within the process of generation of the catalogue of line emitters and further data analysis, we have made use of public Cl0024 catalogues\footnote{\texttt{http://www.astro.caltech.edu/$\sim$smm/clusters/}} from the collaboration ``A Wide Field Survey of Two z=0.5 Galaxy Clusters'' \citep[][hereafter M05]{Treu2003,Moran2005} that include photometric data for 73318 sources detected and extracted in the {\it HST} WFPC2 sparse mosaic covering $0.5\times0.5$\,degrees \citep{Treu2003}  and in ground-based CFHT CFH12k $BVRI$ and Palomar WIRC $JK_s$ imaging.  Visually determined morphological types are given for all sources brighter than I=22.5.  In addition, thousands of photometric and spectroscopic redshift estimates are available. The catalogue of spectroscopically confirmed objects within the field (including foreground, cluster and background sources) comprises 1632 sources \citep[see][and references therein]{Moran2007}.

\section{The Catalogue of ELGs}

\label{sect:catalogue}

In order to produce a catalogue of ELGs we start by selecting the line emitters
The first step to get the catalogue of ELGs is to perform the selection of line emitters
 (either \Ha\ at the redshift of the cluster, or other lines in the case of background contaminants). In many cases, the emission line showed very clear, and even in a fraction of the emitters the \Ha\  and \nii\  lines appeared clearly resolved, but given the number of input sources, an automated or semi-automated procedure was required.  The {\sc tfred} package provides with a task, \texttt{tscale}, that outputs the putative ELGs from the source catalogue, but when applied to our input catalogue, it did not yield reliable results: it was designed for a reduced number of scans and sparse spectral sampling and frequently failed to classify even the most obvious ELGs from our densely sampled pseudo-spectra. Instead, we have followed a different approach, creating  an automatic selection tool, implementing the following steps: \textit{(i)} Define a ``pseudo-continuum'' (hereafter referred as pseudoc) as the subset of pseudo-spectrum points resulting from discarding ``high/low'' outlier values, defined as those above/below the median value $\widetilde{flux}_{alldata}$\,$\pm$\,2\,$\times$\,$\sigma_{alldata}$; the ``pseudo-continuum'' level, $flux_{pseudoc}$, will be defined as its median and the ``pseudo-continuum'' noise, $\sigma_{pseudoc}$, as its standard deviation. \textit{(ii)} The ``upper'' value will be defined as $flux_{pseudoc}$\,+\,2\,$\times$\,$\sigma_{pseudoc}$. Then, the criteria to determine a reliable ELG candidate have been defined as: \textsc{a)} either two consecutive values above ``upper'' are found, or \textsc{b)} or one point above ``upper'' is found and, in addition, one contiguous point above $flux_{pseudoc}$\,+\,$\sigma_{pseudoc}$ and one contiguous point above $flux_{pseudoc}$. These criteria have been chosen since we have observed in our simulations  (see below) that even very narrow lines (0.7\,\AA) always produce a high positive signal in at least two scan slices around the maximum. Single-point peaks in the pseudo-spectrum are attributed to noise or artefacts. On the other hand, the criteria above can cope with broad-line AGNs (see sect. \ref{sect:SF_AGN}).

In order to investigate the reliability of this automatic classifier, we have created a number of simulated spectra comprising an emission line with a Gaussian profile and a flat continuum (a good approximation within our relatively small spectral range). This simple spectrum is convolved with an Airy profile with FWHM\,=\,12\,\AA\ and sampled at steps of 6\,\AA\ to produce a noiseless pseudo-spectrum. Finally, a noise component is built drawing random values from a normal distribution with zero mean and varying standard deviation and added to the pseudo-spectrum signal. We have built a collection of 800 such pseudo-spectra varying different input parameters, namely: {\em (i)} the range of intrinsic line widths of the lines has been adopted from the typical limiting values of the integrated line profiles of giant extragalactic H{\sc ii} regions from \cite{Roy1986}, from some 20\,km s$^{-1}$ to about 40\,km s$^{-1}$ that corresponds to  a range $FWHM_{line}$\,=\,0.7--1.5\,\AA\ (rest frame) with 0.4\,\AA\ step; this range has been further extended by one additional step up to 2.3\,\AA\ (65\,km s$^{-1}$) in order to cope with blending of several H{\sc ii} regions within the galaxy; {\em (ii)} the range of shifts of the peak of the line with respect to the maximum filter transmission has been set to 0--3\,\AA\ with 1\,\AA\ step (i.e. consistent with the scan step of 6\,\AA);  {\em (iii)} the equivalent width (EW) of the emission line has been varied in the range 5--15\,\AA\ with steps of 1\,\AA\ (this range explores our detection sensitivity threshold; at larger EW we do not expect line identification problems) {\em (iv)} and finally, the amplitude (standard deviation) of the added random noise component, $\sigma_{noise}$, has ben set in the range 0.1 to 0.4 times the maximum of the line, with steps of 0.1.

The automatic classifier has identified as line emitters 726 out of the 800 simulated pseudo-spectra (i.e. more than 90\%). From those, only in two cases the classifier chose the source based in a noise feature rather than the correct line. As expected, the 74 pseudo-spectra not classified as ELG are in the high noise range (0.3 or 0.4).    

Moreover, in order to determine the possibility of automatically classifying as ELG a passive galaxy, we have generated in a similar way a set of pseudo-spectra based on a flat continuum and $\sigma_{noise}$ in the range 0.1 to 0.3 times the continuum flux value, with steps of 0.01 (a finer step was chosen in order to create a sufficient large number of instances, given that the noise is the only variable parameter). At each noise step, we have generated 21 instances yielding a total number of 420 pseudo-spectra. From these, 316 (i.e. more than 75\%) have been classifed as passive, while the remaining 104 (i.e. less than 25\%) have been classified as ELG.  

The results of the simulations indicate that our simple classification algorithm is quite effective to classify  true ELGs as a line emitters, but can also pick a non-negligible amount of noisy pseudo-spectra of passive galaxies as emission-line objects. Therefore, we have added an additional step, filtering the sample produced by the selection tool by a careful visual inspection of the pseudo-spectra and also of the thumbnails of all scan slices for every source of this output sample (this was done by three collaborators separately).  After applying these two steps, we have extracted a sample of 210 very robust (i.e. high S/N) ELGs, comprising both star-forming galaxies and AGNs. 

\subsection{Line wavelength estimation}
\label{subsect:line_wave}
Estimating the wavelength of the \Ha\ line is possible with TF tomography, but it is generally a complex issue, since on the one hand, we have a blend of three lines (the \Ha\ line plus the two components of the \nii\ doublet), convolved with the transmission profile of the TF and, as will be described in sect. \ref{sect:ha_flux}, in many cases the pseudo-spectrum line ``profile'' (hereafter referred as line pseudo-profile) is affected by absorption-like features. We have attempted to derive the \Ha\ line position  considering a model comprising three Gaussian lines plus a linear continuum; the rest-frame wavelength relative positions are fixed, as is the ratio of the two \nii\ doublet components (set to $f_{6548}/f_{6583} = 0.3$). Free parameters of the model are: the observed wavelength of the \Ha\ line, the line width (constrained to be the same for the three lines), the \nii\ $\lambda$\,6583 and \Ha\ fluxes and the continuum level. This model spectrum has been convolved with the TF transmission profile and fitted by means of non-linear least squares to the pseudo-spectra profiles. For some 30\% of the sources, the result of the fit reproduces accurately the pseudo-spectrum profile, but in a vast majority of the cases the fit either fails or provides inaccurate results due to noise in the pseudo-continuum, absorption-like features in the line pseudo-profile, etc. Eventually, we have decided to derive the position of the line by manually fitting the pseudo-spectrum using the IRAF \texttt{splot} task and either a Gaussian or a Lorentzian profile, choosing after inspection the appropriate range to avoid continuum noise and contaminant lines.  There is a very good agreement between the line positions computed by \texttt{splot} and those resulting from the trustful, accurate model fits ($\sim$\,1\,\AA). In a minority of the cases, where the line profile showed very asymmetric (e.g. when absorption-like features are present, most likely due to random noise as shown in sect. \ref{sect:ha_flux} below), the position of the line was chosen to be the peak value of the pseudo-spectrum. Given the difficulty of providing a trustful uncertainty figure, we have assumed a constant error value of 3\,\AA\ for the fit to the peak of the line, i.e. half of a scan step; this error is square-added to the tuning uncertainty of 1\,\AA\ and to the wavelength error introduced by the combination of images performed in order to produce the pseudo-spectrum (see sect. \ref{TFreduction}), also considered to be 1\,\AA\ at most; hence,  $\sigma_{pos}$\,$\simeq$\,3.3\,\AA.

\subsection{Identification of \Ha\ emitters: rejection of interlopers}
\label{subsect:outliers}
\begin{figure*}[tb]
   \centering
   %\vspace{-2cm}
   %\hspace{-1.5cm}
   \includegraphics[scale=0.55]{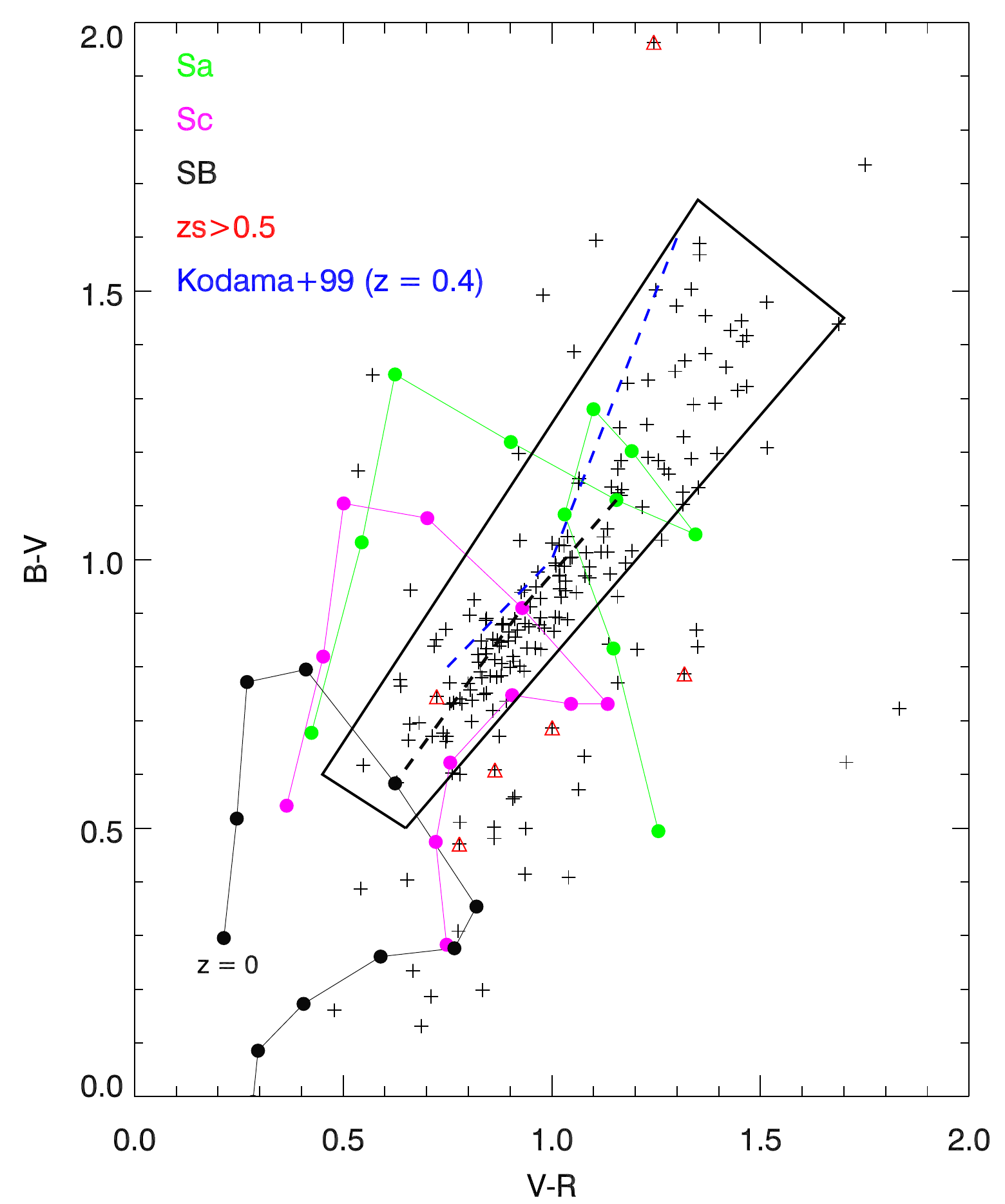}
   %\hspace{-1.5cm}
   \includegraphics[scale=0.55]{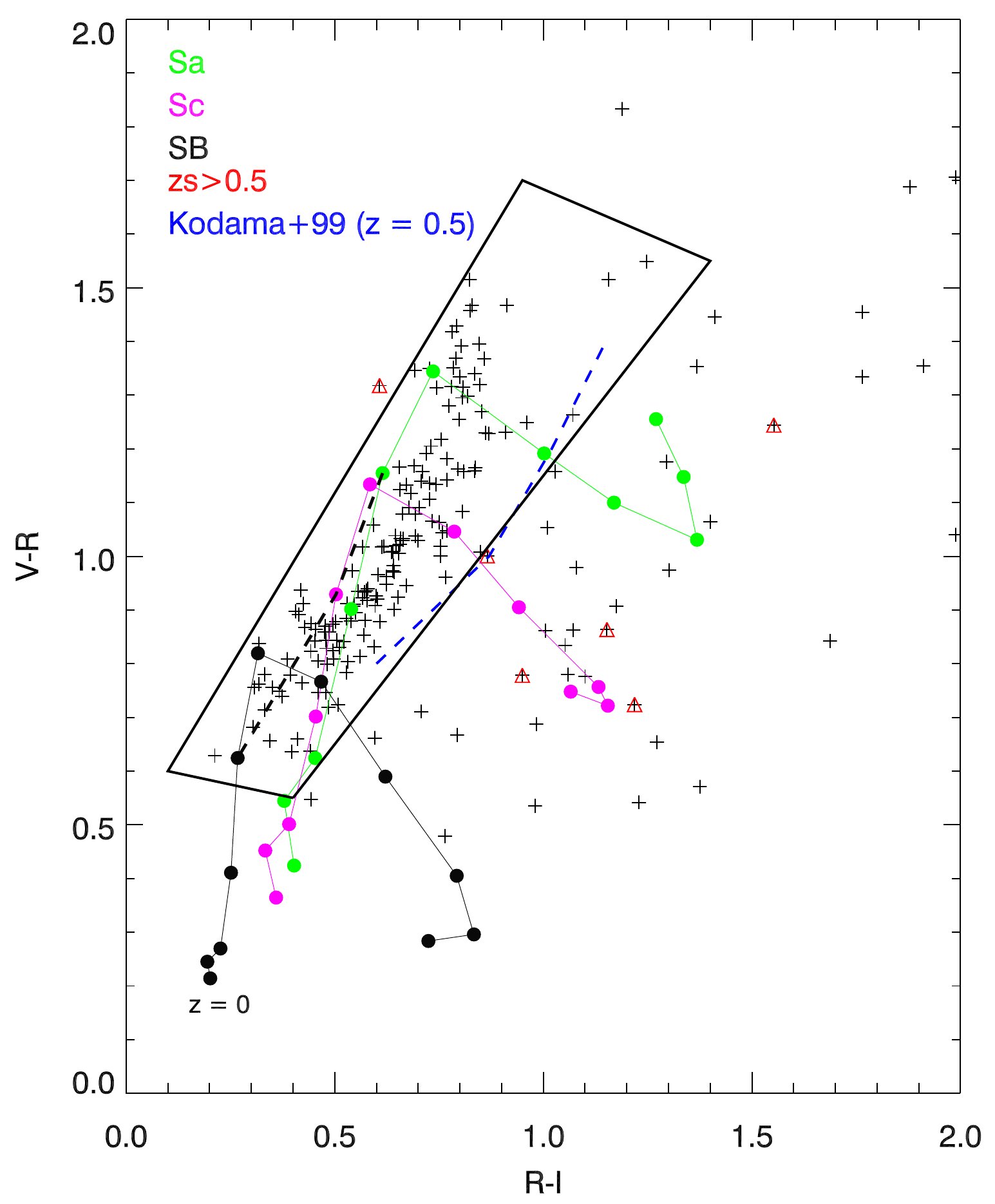}
      \caption{\mbox{$B-V$} vs. \mbox{$V-R$} (left) and \mbox{$V-R$} vs. \mbox{$R-I$} (right) diagnostic diagrams. The sample, comprising 202 unique objects with counterparts in the M05 catalogue is denoted by cross symbols ($+$). The red triangles ($\triangle$) mark objects with spectroscopic redshift $z_s > 0.5$ in the M05 catalogue.  The control grid has been built using synthetic colours from COSMOS templates for Sa, Sc and SB galaxies in the range z\,=\,0.0--1.0 with steps of $\Delta$z\,=\,0.1. The thick, black dashed line connects the three templates at z\,=\,0.4. The blue dashed line depicts the \cite{Kodama1999} models at z\,=\,0.4 and z\,=\,0.5. The thick, black solid line box marks the approximate region used for selecting the cluster candidates.
              }
         \label{fig:interlopers}
   \end{figure*}

The observations presented here target a single emission line; therefore, a number of interlopers can be present; these are expected to be ELGs at other redshifts. When observing \Ha\  at z\,=\,0.40, 
\oiii$\lambda$\,5007\,\AA\ emitters at z\,=\,0.83 and \oii$\lambda$\,3727\,\AA\  
emitters at z\,=\,1.46 can be detected as well. However, at the 
limiting fluxes considered here these contaminants are not expected to be very abundant and can be easily discriminated via colour-colour diagrammes. We have implemented such colour-colour diagnostic diagrams following \cite{Kodama2004} according to the following steps: first, we have matched our initial catalogue of 210 robust candidates with the photometric and spectroscopic catalogue from M05 and a matching radius of 1.0\,arcsec (i.e. compatible with the accuracy of our astrometry). A counterpart  with at least photometric data has been found for 202 sources. Using the CFHT CFH12k aperture photometry, we have computed the \mbox{$B-V$}, \mbox{$V-R$} and \mbox{$R-I$} colours. The diagnose grid has been built by deriving synthetic colours for COSMOS templates \cite{Ilbert2009} of Sa, Sc, and starburst (SB) galaxies at redshifts ranging from z\,=\,0 to z\,=\,1.0 with steps of 0.1, integrated to the CFH12k $B$, $V$, $R$ and $I$ passbands. Also, models from from \cite{Kodama1999} at z\,=\,0.4 and z\,=\,0.5 have been included in the grid. The diagrammes for  \mbox{$B-V$} vs. \mbox{$V-R$} and \mbox{$V-R$} vs. \mbox{$R-I$} are shown in Fig. \ref{fig:interlopers}. %\textbf{falta marcar la regi\'on en los diagramas}. 
A source has been deemed as interloper if: \textit{(i)} the source is outside the cluster region (as depicted in Fig.  \ref{fig:interlopers}) in both colour diagrammes and does not have a spectroscopic redshift from M05 within the cluster range  (set as 0.35\,$\leq$\,z\,$\leq$\,0.45); \textit{(ii)} the source has a spectroscopic redshift from M05 outside the cluster range, independent of its colours. 

\begin{figure*}[!t]
   \centering
   %\vspace{-2.5cm}
   %\hspace{-1.5cm}
   \includegraphics[scale=0.85]{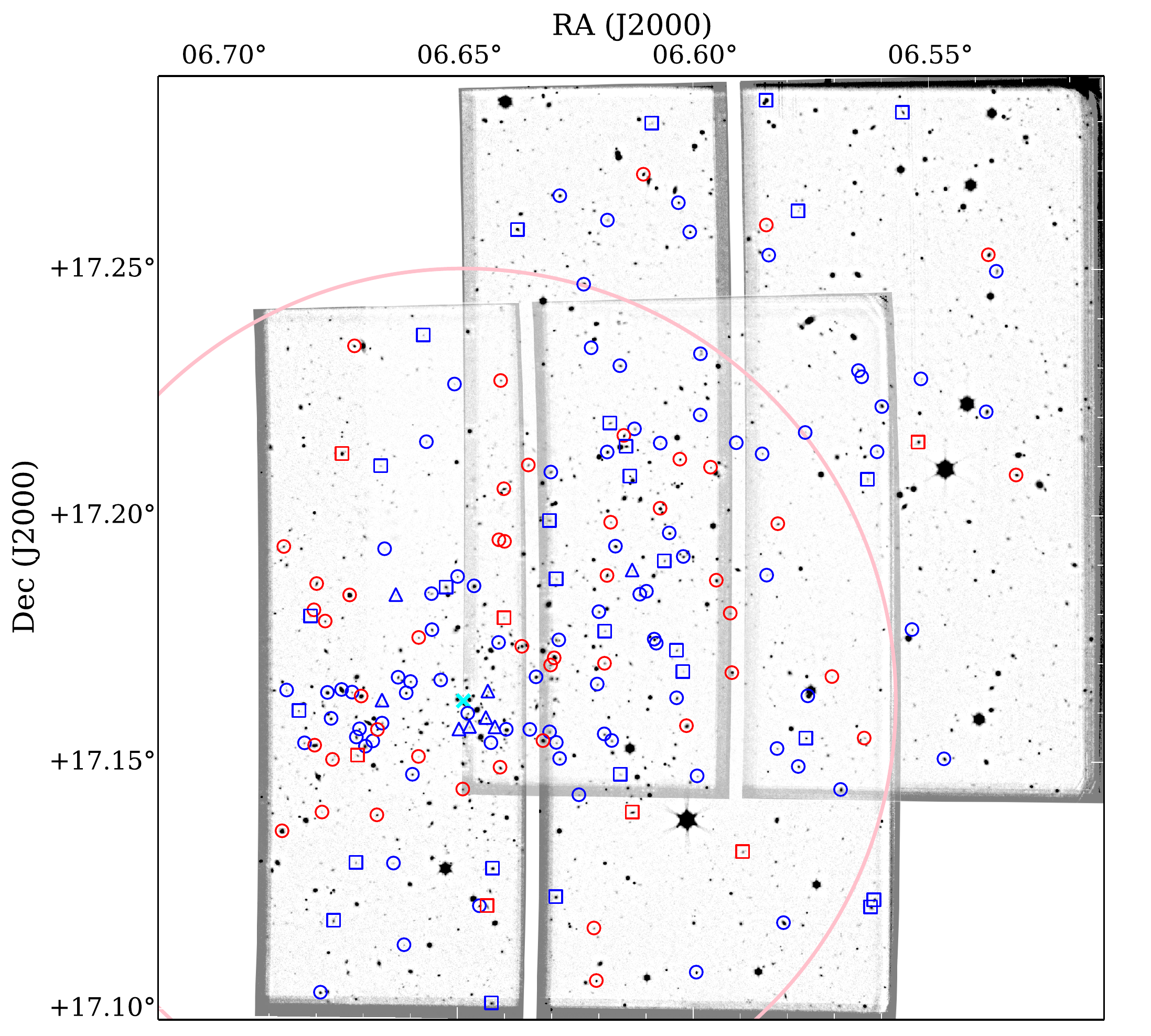}
   %\vspace{-2.5cm}
    \caption{
The figure depicts the  two OSIRIS/GTC pointings towards Cl0024. The plot has standard orientation: north is at the top and east to the left. Blue symbols correspond to SF galaxies  and red ones correspond to AGNs according to the \niiHa\,$\geq$\,0.6 criterion from \cite{Ho1997}. Circles correspond to galaxies in the main cluster structure (structure ``A'' in sect.~\ref{sect:redshift}), squares to those in the infalling group (structure ``B'') and triangles to sources in the putative group around z\,$\approx$\,0.42 (see  sect.~\ref{sect:redshift}). The cyan cross ($\times$) denotes the centre of the cluster (galaxies/BCG) and the large pink circle marks the virial radius of 1.7 Mpc \citep{Treu2003}. A general alignment of ELGs is observed in the NW – SE direction, consistent with a a structure assembling onto the cluster core from the NW with an orientation almost in the plane of the sky \citep{Moran2007,Zhang2005,Kneib2003}.     
      }
\label{fig:map_and_sources} 
   \end{figure*}

Based on the first criterion, we have deemed 19 sources as contaminants. These are in a vast majority of the cases located in the locus of the colour-colour diagram occupied by galaxies at  z\,=\,0.8--1.0, i.e. consistent with being \oiii$\lambda$\,5007\,\AA\ emitters. Moreover, 97\% of the ELGs with a redshift from M05 within the cluster range have been also catalogued as cluster members by the colour-colour diagnostic described above. The second criterion added 9 additional sources as interlopers (two sources have been discarded by both colours and redshift criteria). Hence, 28 objects have been classified as contaminants at a different redshift. In addition, 8 sources do not have a counterpart in the M05 catalogue and therefore have been excluded from the cluster list. Therefore, our final catalogue consists of 174 robust cluster \Ha\ emitters (see the distribution of sources in the sky in Fig. \ref{fig:map_and_sources}), 28 putative interlopers (ELGs at a different redshift, mostly oxygen emitters at z\,$\sim$\,0.9) and 8 sources without ancillary data to perform the assessment. Future analysis of additional spectral ranges (centred at the \oiii\ and \Hb\ wavelengths at the nominal redshift of Cl0024) will refine this rejection criterion. The 174 sources in our final catalogue of unique robust cluster emitters are listed in Tab.~\ref{tab:cldata}. From these, 112 have spectroscopic redshifts in the M05 catalogue. 

\begin{figure*}[!t]
   \centering
   %\vspace{-1cm}
   \hspace{-1cm}
   \includegraphics[scale=0.37]{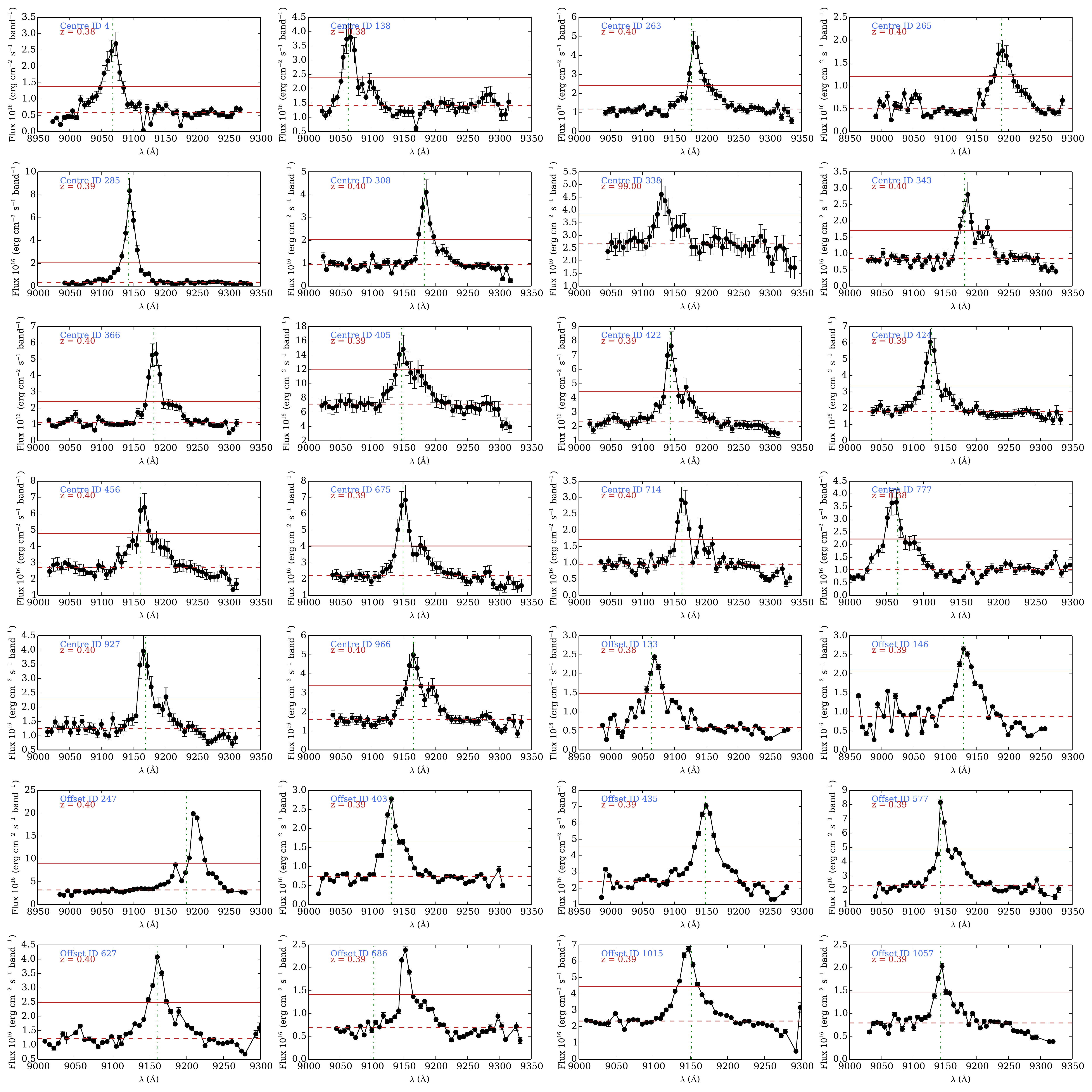}
      \caption{Selection of 28 pseudo-spectra with high signal-to-noise. The green dashed-dotted lines correspond to spectroscopic or secure photometric redshifts from M05 (when available). The dashed red line corresponds to the approximate pseudo-spectrum continuum level, and the solid red line to the 3$\sigma_{cont}$ level, where $\sigma_{cont}$ is the pseudo-continuum noise. For the source ID~offset/686, the redshift from M05 is photometric. 
              }
         \label{fig:pseudo_highsn}
   \end{figure*}

Fig. \ref{fig:pseudo_highsn} shows a selection of high signal-to-noise pseudo-spectra. In many cases, the strongest \nii\ doublet component (at 6583\,\AA) is clearly separated from the \Ha\ line in a visual inspection. Sometimes the line is observed as a ``shoulder'' at the long wavelength side of  the \Ha\ line. 

\section{Derivation of line fluxes}

\label{sect:ha_flux}

From the pseudo-spectra described in sect. \ref{sect:catalogue}, it is possible to derive the \Ha\ and \nii\ fluxes following several approaches. We have applied a straightforward procedure derived from the standard narrow-band on-band/off-band technique, using for each source the flux in the scan slice closest to the computed position of the \Ha\ line (see sect. \ref{sect:catalogue}) and that of the slice closest to the \nii\ line; as will be shown below, this method, though simple, produces acceptable results when compared to the more sophisticated procedure based on least-squares fitting  of the pseudo-spectrum to a model spectrum convolved with the transmission profile of the TF described in sect. \ref{subsect:line_wave} with the advantage that the former method is always applicable while the latter can be only used in a minority of cases. 

We start by subtracting a linear continuum. This can be easily done by applying a linear fit to the regions of the pseudo-spectrum excluding the emission line. Then, assuming infinitely thin lines, the  \Ha\ and  \nii\ line fluxes, denoted by $f\left(H\alpha\right)$ and $f\left(\left[\ion{N}{ii}\right]\right)$ respectively, are  given by the expressions (Cepa, priv. comm.):

\begin{eqnarray}
\label{eq:fluxha_n2}
f_{on,H\alpha} & = & T_{H\alpha}\left(H\alpha\right) f\left(H\alpha\right)  +  T_{H\alpha}\left(\left[\ion{N}{ii}\right]\right)f\left(\left[\ion{N}{ii}\right]\right)\nonumber\\
f_{on,\left[\ion{N}{ii}\right]} & = & T_{\left[\ion{N}{ii}\right]}\left(H\alpha\right) f\left(H\alpha\right)  +  T_{\left[\ion{N}{ii}\right]}\left(\left[\ion{N}{ii}\right]\right) f\left(\left[\ion{N}{ii}\right]\right)
\end{eqnarray}

\noindent where $f_{on,H\alpha}$ and $f_{on,\left[\ion{N}{ii}\right]}$ are the continuum-subtracted  fluxes in the chosen \Ha\ and \nii\ slices and $T_{<slice>}\left(<line>\right)$ denotes the TF transmission of a given slice at a given line wavelength. The different transmission values can be easily derived from the approximate expression given in Eq. \ref{eq:transmission}. From Eq.~\ref{eq:fluxha_n2} we can easily derive the flux in the \Ha\ line:

\begin{equation}
\label{eq:flux_ha}
f\left(H\alpha\right) = \frac{f_{on,H\alpha} T_{\left[\ion{N}{ii}\right]}\left(\left[\ion{N}{ii}\right]\right) -  f_{on,\left[\ion{N}{ii}\right]} T_{H\alpha}\left(\left[\ion{N}{ii}\right]\right)}{T_{H\alpha}\left(H\alpha\right) T_{\left[\ion{N}{ii}\right]}\left(\left[\ion{N}{ii}\right]\right) - T_{H\alpha}\left(\left[\ion{N}{ii}\right]\right) T_{\left[\ion{N}{ii}\right]}\left(H\alpha\right)}
\end{equation}

And a similar expresion for the \nii\ line. The errors in the lines have been derived by propagation, taking into account not only the errors in the \Ha\ and \nii\ ``on'' bands, but also the continuum noise (i.e. the noise around the zero-level continuum after removing the linear fit explained above). Hence, the line error has been computed as:

\begin{eqnarray}
\label{eq:error_flux_ha}
\Delta f\left(H\alpha\right) &  = & (\left(T_{\left[\ion{N}{ii}\right]}\left(\left[\ion{N}{ii}\right]\right) \Delta f_{on,H\alpha}\right)^2 \\
& &   +  \left(T_{H\alpha}\left(\left[\ion{N}{ii}\right]\right) \Delta f_{on,\left[\ion{N}{ii}\right]}\right)^2 ) \nonumber\\
& & +  \left(\left(T_{\left[\ion{N}{ii}\right]}\left(\left[\ion{N}{ii}\right]\right) - T_{H\alpha}\left(\left[\ion{N}{ii}\right]\right)\right)\sigma_{cont}\right)^2)^{1/2}\nonumber\\
& & / \left(T_{H\alpha}\left(H\alpha\right) T_{\left[\ion{N}{ii}\right]}\left(\left[\ion{N}{ii}\right]\right) - T_{H\alpha}\left(\left[\ion{N}{ii}\right]\right) T_{\left[\ion{N}{ii}\right]}\left(H\alpha\right)\right)\nonumber
\end{eqnarray}

\noindent where $\Delta f_{on,H\alpha}$ and $\Delta f_{on,\left[\ion{N}{ii}\right]}$ are the flux errors in the ``on'' \Ha\ and \nii\ bands computed as indicated in sect.~\ref{subsect:flux_calib} and $\sigma_{cont}$ is the continuum error measured as the standard deviation of the points within the region of the pseudo-spectrum excluding the emission lines. The contribution of the continuum noise to the total error is important, on average $\sim$\,30\% at the central position and much larger at the offset position where the exposure times are smaller, on average 60--70\%.

The median fractional error in the \Ha\ fluxes is $\approx$\,24\%. A 70\% of the sample objects have relative errors below 30\%. This errors are compatible with those quoted by \cite{Lara2010} (see sect.~\ref{subsect:implementation}), though somewhat larger than those derived from their simulations due to our larger continuum errors. However, since the \nii\ line is usually fainter than the \Ha\ line, its flux errors are in general notably larger: the average fractional error is $\approx$\,54\% and only 15\% of the sample objects have a relative error below 30\%. This was of course expected since the detection/selection algorithm is driven by the strongest line present in the pseudo-spectrum. Hence, in many cases the \Ha\ line acts as a ``prior'' and the nitrogen flux is extracted at the expected wavelength of the (otherwise barely detected) \nii\ line.

As mentioned above, the line flux estimation is based in an infinitely thin line approximation which assumes that the line can be well represented by $\delta(\lambda - \lambda_z)$, where $\lambda_z$\,=\,$\lambda_0$(1 + z). According to \cite{Pascual2007}, for star-forming galaxies, emission line widths are mass-related and typically FWHM\,$\lesssim$\,10\,\AA\,$\times$\,(1 + z), and, for narrow-band filters of some 50\,\AA\ width, it is possible to recover $\sim$\,80\% of the line flux up to z\,$\sim$\,4. We have investigated the impact of applying such approximation to our very narrow TF scans ($\sim$\,12\,\AA). To this end, we have performed several simulations using Gaussian line profiles of several widths peaking at different offsets with respect to the maximum of the filter transmission profile (Eq. \ref{eq:transmission}). The emission line broadening is given by the relation \citep{FernandezLorenzo2009}:

\begin{equation}
\label{eq:line_width_Vmax}
2V_{max} = \frac{\Delta\lambda c}{\lambda_0 \sin(i) (1 +z)}
\end{equation}

\noindent where V$_{max}$ is the maximum rotation velocity, $\lambda_0$ is the line wavelength at z\,=\,0 and $\Delta\lambda$ is the line width at 20\% of peak intensity. For a Gaussian line, $\Delta\lambda$\,=\,1.524\,$\times$\,FWHM. Assuming V$_{max}$\,=\,200\,km\,s$^{-1}$ as a safe upper limit \citep[see for instance][]{FernandezLorenzo2009}, the \Ha\ line FWHM\,$\simeq$\,8\,\AA\ (here we assume that the line is unresolved. The possibility that the line appears  resolved in our pseudo-spectra due to kinematical split is investigated below). For narrow lines, FWHM\,=\,2\,\AA, we are able to recover 96--98\% of the flux (from 0 to 2\,\AA\ offset), while for the widest simulated lines, FWHM\,=\,8\,\AA, the fraction of recovered flux is in the range 73--76\%. In order to compare the simulations  with real results, we have used the small set of pseudo-spectra for which a reliable fit to the model spectrum was achieved, finding that the average ratio between the flux derived from the infinitely thin line approximation and that derived from the best fit is 0.81 and  0.88 for the \Ha\ and \nii\ lines, respectively, i.e. well aligned with the results of our simulations.

\begin{figure*}[tb]
   \centering
   %\vspace{-1cm}
   \hspace{-2cm}
   \includegraphics[scale=0.4]{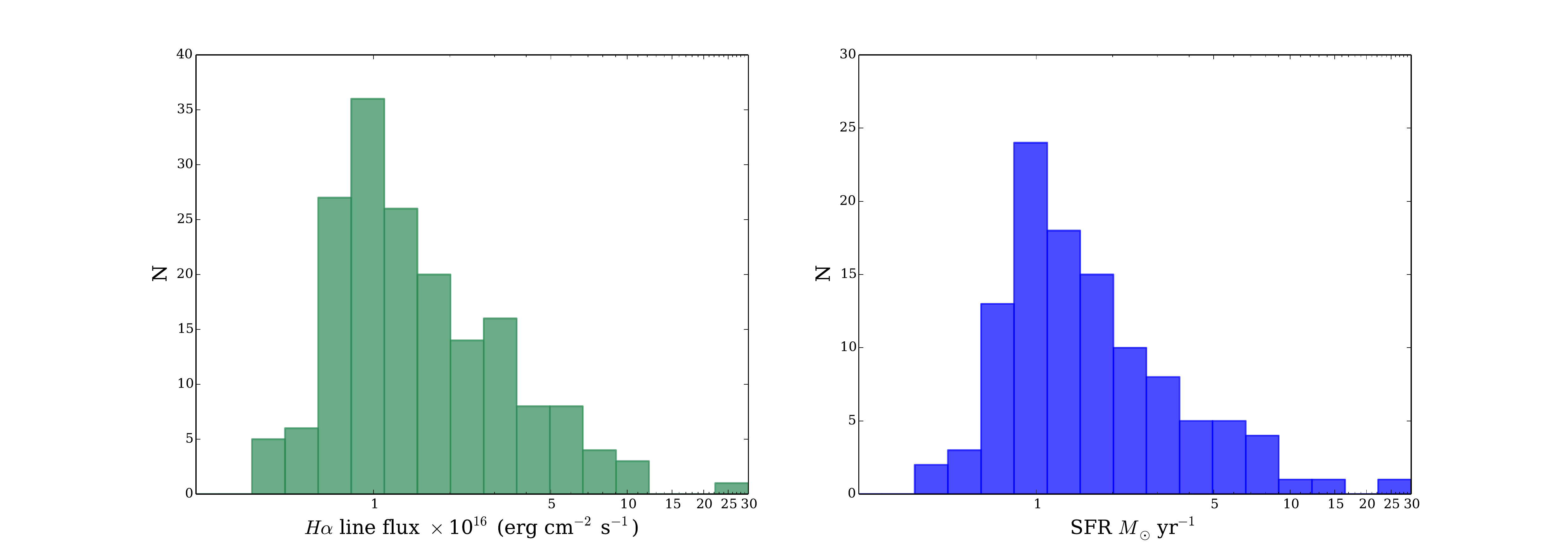}
      \caption{\textit{Left}: histogram of \Ha\ line fluxes for the 174 robust cluster members (after correcting for the strongest component of the \nii\ doublet). \textit{Right}: histogram of SFR for SF galaxies (i.e. excluding AGN candidates) from the \Ha\ fluxes, applying a standard 1\,mag. extinction correction for the line and the \cite{Kennicutt1998} luminosity-SFR conversion.}
         \label{fig:flux_histo}
   \end{figure*}

The completeness limit of the ELG sample, as given by the maximum of the flux histogram depicted in Fig. \ref{fig:flux_histo} (left panel) is $\sim$\,0.9\,$\times$\,10$^{-16}$\,erg\,s$^{-1}$cm$^{-2}$, i.e. better than the GLACE requirements.

% section 4.2.3. - Flux correction: underlying stellar absorption vs. rotational splitting. %

\subsection{Absorption-like features in the pseudo-spectra}

\label{subsect:absorption_like}

\begin{figure*}[!t]
   \centering
   %\vspace{-1cm}
   \hspace{-1cm}
   \includegraphics[scale=0.37]{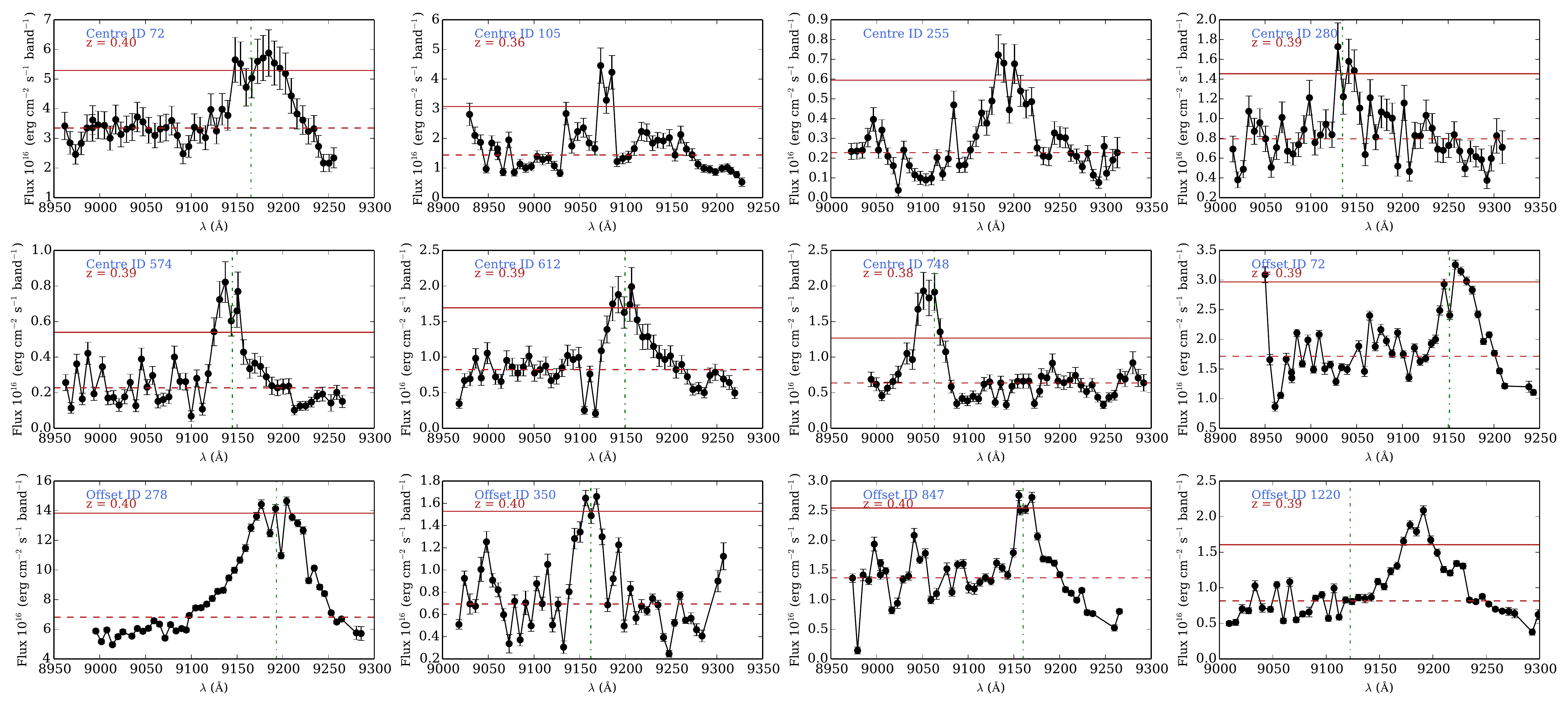}
      \caption{Selection of pseudo-spectra with a variety of absorption-like features. Lines as in Fig. \ref{fig:pseudo_highsn}. For the source ID~offset/1220, the redshift from M05 is photometric. 
              }
         \label{fig:pseudo_absorption_like}
   \end{figure*}

In a large number of cases, we have observed  in the pseudo-spectra absorption-like features affecting the (putative) \Ha\ emission line.  This affects around 50\% of the galaxies, and shows very clear in about 25\% of the cases, as those depicted in Fig.~\ref{fig:pseudo_absorption_like}. We have explored a number of possible explanations: first, and perhaps most obvious, that the absorption-like features are due to random noise. These characteristics have been observed in the simulations described in the preceding sections, in  approximately 20\% of the cases (some 146 out of 726 simulated pseudo-spectra classified as ELGs). This is therefore a very probable explanation in an ample fraction of cases. In addition, we have explored two potential physical explanations: either an emission line split due to the galaxy rotation or the presence of underlying stellar absorption.  In order to investigate the first possible cause, we have performed a series of simulations similar to those described in the preceding section but starting with   a spectrum comprising two identical Gaussian lines with separation $2\lambda_0(1+z)V_{max}/c$ and a flat continuum in order to re-create the effect of the rotational split in TF pseudo-spectra. We have built our  
collection of pseudo-spectra varying the range of intrinsic line widths in the range $FWHM_{line}$\,=\,0.7--2.3\,\AA\ (rest frame) with 0.4\,\AA\ step; the range of shifts of the peak of the line with respect to the maximum filter transmission has been set to 0--3\,\AA\ with 1\,\AA\ step and finally the  galaxy rotational speed has been let to vary in the range $V_{max}$\,=\,100--200\,km\,s$^{-1}$ with 10\,km\,s$^{-1}$ step, in agreement with  \cite{FernandezLorenzo2009}; hence, the separation between the two line peaks ranges from some 6.1\,\AA\ to about 12.3\,\AA; and finally the amplitude of the added random noise component ($\sigma_{noise}$) has ben set in the range 0.1 to 0.3  times the maximum of the line, with steps of 0.1.

After inspecting 1320 pseudo-spectra simulated as described above, we have observed that the fraction of absorption-like features observed in the pseudo-spectra is relatively low at rotational speeds $\leq$\,160\,km s$^{-1}$, ranging from approximately 10\% at 100\,km s$^{-1}$ to some 26\% at 160\,km\,s$^{-1}$ and therefore consistent with noise-induced features. However, the fraction of features observed raises notably for $V_{max}$\,$\geq$\,170\,km s$^{-1}$, being of 45\% at 170\,km\,s$^{-1}$ and 58\%--60\% above that speed. Therefore, a kinematical line split due to galaxy rotation is a plausible mechanism to induce a fraction of the absorption-like features observed. Nevertheless, it should be taken into account that the fraction of such fast-rotating galaxies is reduced \citep[around 14\% in the sample of ][in the range 0.3\,$<$\,z\,$<$\,0.8]{FernandezLorenzo2009} and moreover that we are assuming $\sin i = 1$, i.e. edge-on or nearly edge-on galaxies so the actual fraction of objects fulfilling the conditions is probably on the order of 14\% at most. 

We have finally explored the possibility that these absorption-like features are caused by true absorption due to the underlying host galaxy stellar component. To this end, the most likely host galaxy stellar population was derived by fitting the public photometric information from  M05 (B, V, R, I from CFHT; J, K$_s$ from WIRC at Palomar 200\textquotedblright) by means of the \textit{LePhare} code \citep{Ilbert2006} using the set of  SED templates from \cite{Bruzual2003}, star formation histories exponentially declining with time as $\mathit{SFR} \propto e^{-t/\tau}$ with $\tau$ ranging from 0.1 to 30.0 Gyr, initial mass function (IMF) from \cite{Chabrier2003} and dust extinction law from \cite{Calzetti2000}. The details of the computations will be thoroughly described in P\'erez-Mart´\'{\i}nez et al. (in prep.). 
%but the overall procedure is summarised below: 
%the set of  SED templates from \cite{Bruzual2003} was used, with star formation histories exponentially declining with time as $SFR \propto e^{-t/\tau}$. Nine different values of $\tau$ (0.1, 0.3, 1.0,
%2.0, 3.0, 5.0, 10.0, 15.0 and 30.0 Gyr) with 221 steps in age and three different metallicity values (42, 52, 62) were applied. The used IMF is that of \cite{Chabrier2003}. Dust extinction  was applied in the SED-fitting procedure using the \citet{Calzetti2000} extinction law, with a maximum E(B-V) value of 1.4. A condition imposed was that the derived age of the galaxies must be less than the age of the Universe at that redshift and  greater
%than $10^{8}$ years (the latter requirement avoids having galaxies with extremely
%high specific star formation rates, $sSFR=SFR/M$. The output of the procedure included the best-fitting host galaxy stellar population template, the stellar mass, age and extinction. 
%%We estimated the 1$\sigma$ uncertainty of the galaxy stellar mass as half the difference between $M_{sup}$ and $M_{inf}$, where  $M_{sup}$ and $M_{inf}$ are the mass values obtained for $\Delta \chi^2 = 1$, given as an output of the \textit{LePhare} code. The mean error on the galaxy stellar mass, as estimated from  the $1\sigma$ dispersion of the mass distribution, is 0.08 dex. 
The best-fitting host galaxy stellar template was furthermore re-sampled at a finer resolution using GALAXEV\footnote{\texttt{http://www2.iap.fr/users/charlot/bc2003/}} \citep{Bruzual2003} and then convolved with the TF transmission profile to obtain a ``host galaxy stellar component'' pseudo-spectrum. The pseudo-continuum of this synthetic pseudo-spectrum was then scaled to that of the real object within the range of the observations. In some cases, the absorption features were still present in the synthetic pseudo-spectrum and could account for at least a fraction of the absorption-like feature observed. However, due to the nature of the procedure, the best fit stellar template solution is notably degenerate when taking into account the photometric errors and therefore the flux correction is uncertain. Hence, we have decided to omit such correction for the \Hanii\ line fluxes. 

\section{Spatial and redshift distribution of the cluster galaxies}
\label{sect:redshift}

Spectroscopic redshifts from M05 were available for 112 of our ELGs (see sect.~\ref{subsect:outliers}). In all cases, the M05 redshifts  also placed our cluster candidates within the cluster.  There was a remarkably good agreement between the redshift estimates derived from the position of the \Ha\ line within our pseudo-spectra (see line positions derived from spectroscopic redshifts in Fig. \ref{fig:pseudo_highsn}) and that derived from spectroscopic measurements. In fact, the redshift error defined as $\vert z_{TF} - z_{spec} \vert / ( 1 + z_{spec}) $ is on average 0.002 (median value 0.0005) with a maximum value of 0.02, i.e. we can consider the TF-derived redshifts as of spectroscopic quality. 

\begin{figure}[htb]
   \centering
   %\vspace{-7cm}
   %\hspace{-0.2cm}
   \includegraphics[scale=0.27]{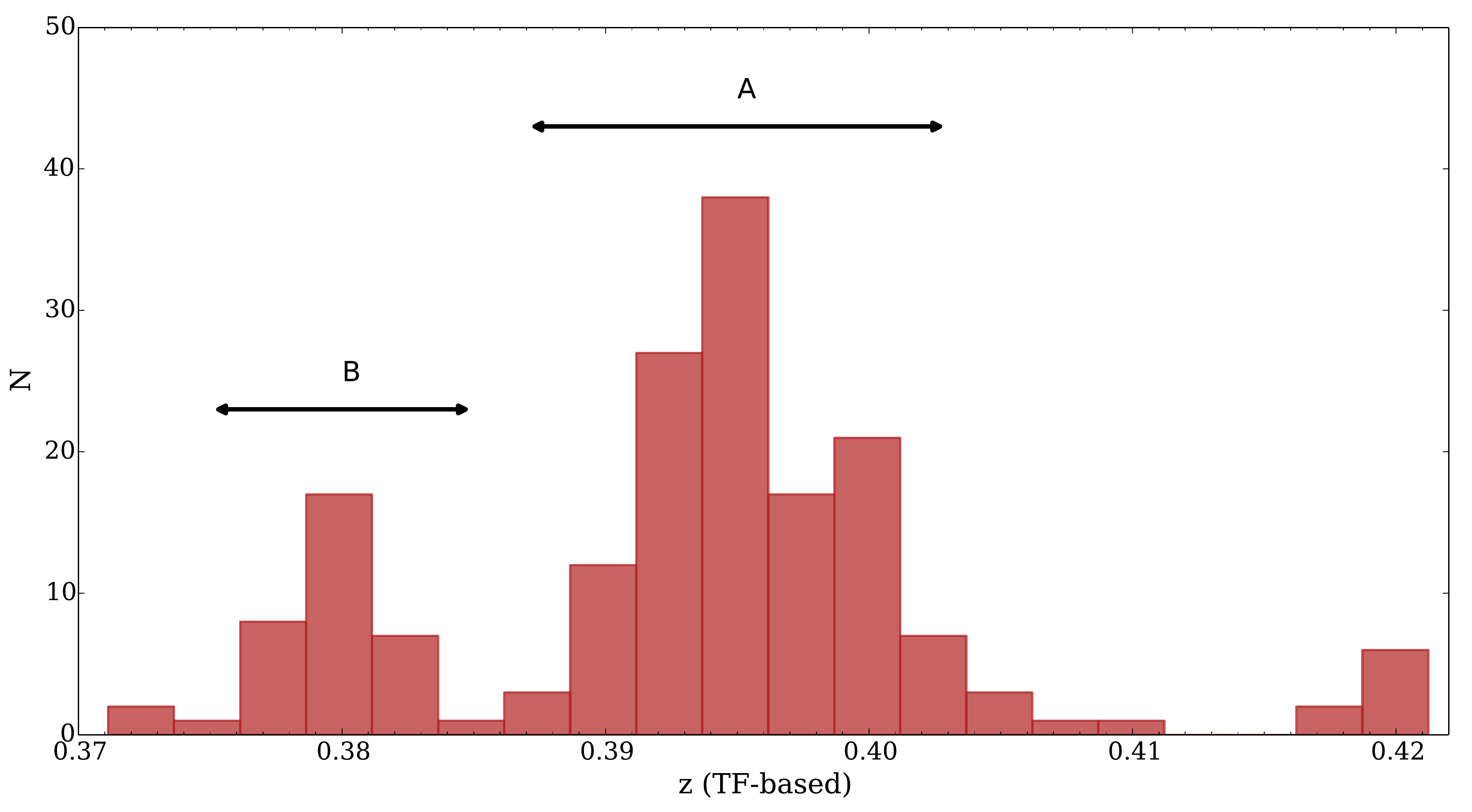}
       \caption{ Distribution of redshifts derived from our data. It is possible to recognize two dynamical structures as in \cite{Czoske2002}: ``A'' is the main cluster component, while ``B'' lies along the line of sight to the cluster centre and has been interpreted as an infalling group at high velocity. 
              }
         \label{fig:redshift_distrib}
   \end{figure}

The spatial distribution of the ELG sample, as given by the position of the sources in the sky plane and our redshift estimates, maps the presence of two components: {\it (i)} a structure assembling onto the cluster core from the NW with an orientation almost in the plane of the sky. This structure has been already reported by other authors \citep{Moran2007,Zhang2005,Kneib2003}. {\it (ii)} An infalling group at high velocity nearly along the line of sight to the cluster centre, identified by a double-peaked distribution in the redshift space, as shown in Fig. \ref{fig:redshift_distrib} \citep{Moran2007,Czoske2002}: structure ``A'' (centred at z\,=\,0.395) and ``B'' (centred at z\,=\,0.381) correspond to the main cluster and infalling group components, respectively.

\begin{figure}[htb]
   %\centering
   \hspace{-0.5cm}
   \includegraphics[width=1.15\columnwidth]{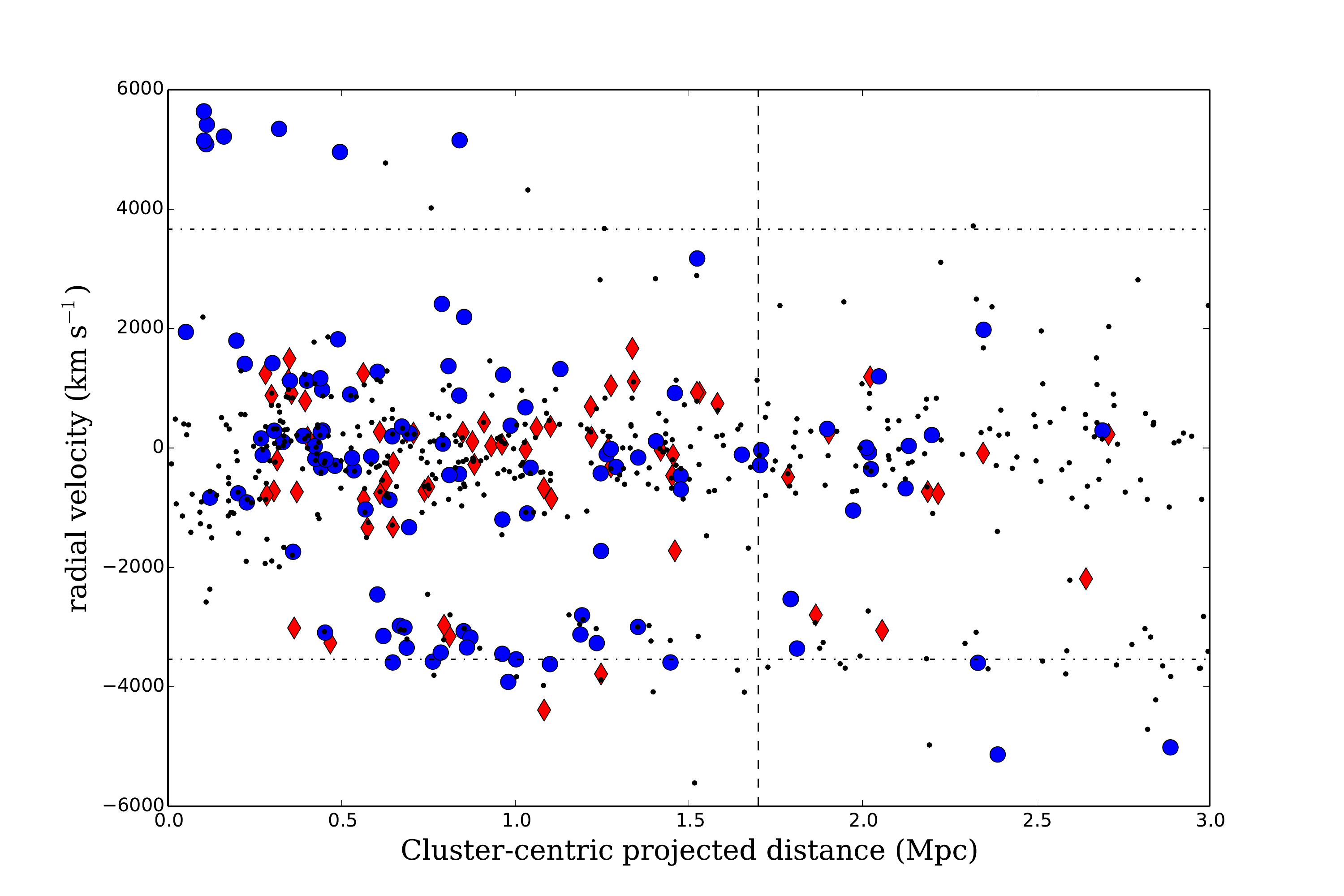}
      \caption{Line-of-sight velocity relative to the cluster main component central redshift z\,=\,0.395 plotted against the projected distance to the cluster centre (as defined by the distribution of galaxies/BCG). Red diamonds correspond to AGNs and blue dots to SF galaxies according to the H97 criterion (see sect.~\ref{sect:SF_AGN}). The dotted vertical line marks the virial radius of 1.7 Mpc \citep{Treu2003}. The dashed-dotted horizontal lines correspond to the radial velocity limits fully covered within the field of view of both OSIRIS TF pointings (see sect.~\ref{sect:obs_data_reduction} and Eq.~\ref{radial_dep}). The small, black points correspond to galaxies from M05 with spectroscopic redshifts, including both passive and star-forming objects.}
         \label{fig:elg_vrad_rclus}
   \end{figure}

The distribution of galaxies can be also seen in Fig.~\ref{fig:elg_vrad_rclus}, where the radial velocities relative to the central redshift of component ``A'' are plotted against the distance to the cluster centre as given by the distribution of galaxies/BGC \citep{Treu2003}. The two mentioned components are clearly separated. There is a third, small group of 8 galaxies at z\,$\approx$\,0.41--0.42 (clearly seen in both  Fig.~\ref{fig:elg_vrad_rclus} and  Fig.~\ref{fig:redshift_distrib}) that has been also reported by \cite{Czoske2002}. While according to these authors these galaxies are most likely part of the surrounding field galaxy population, we cannot rule out the possibility of another group connected to Cl0024. This suggestion could be reinforced by the fact that all these galaxies are observed within a relatively small cluster-centric projected distance, r\,$<$\,1\,Mpc. While this can be just a instrumental effect, result of the incomplete wavelength coverage at the red side of the velocity field, is should be taken into account that, on the one hand, the cluster centre is offset some 1\,arcmin to the SE of the TF axis of the central pointing, and on the other, that no other excess of ELG is observed close to the TF axis of the offset pointing, as would be expected if the galaxies were part of the field galaxy population.

%smaller than the systematic uncertanties mentioned avove.

\section{\Ha\,$+$\,\nii\ luminosity function}
\label{sect:Ha_LF}

\begin{figure*}[htb]
   %\centering
   %\vspace{-7cm}
   %\hspace{-1.5cm}
   \includegraphics[scale=0.5]{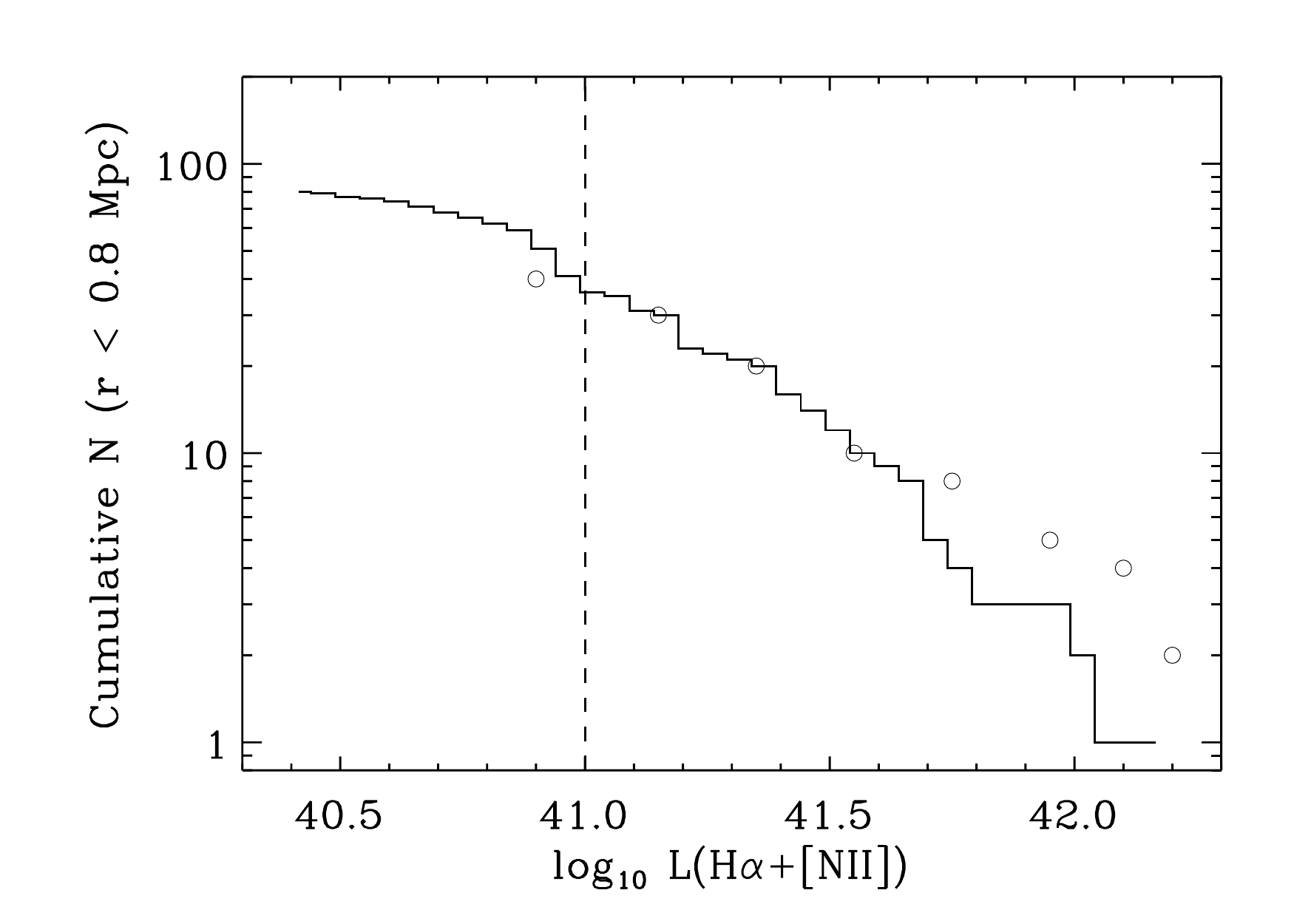}
   %\hspace{-1.5cm}
   \includegraphics[scale=0.5]{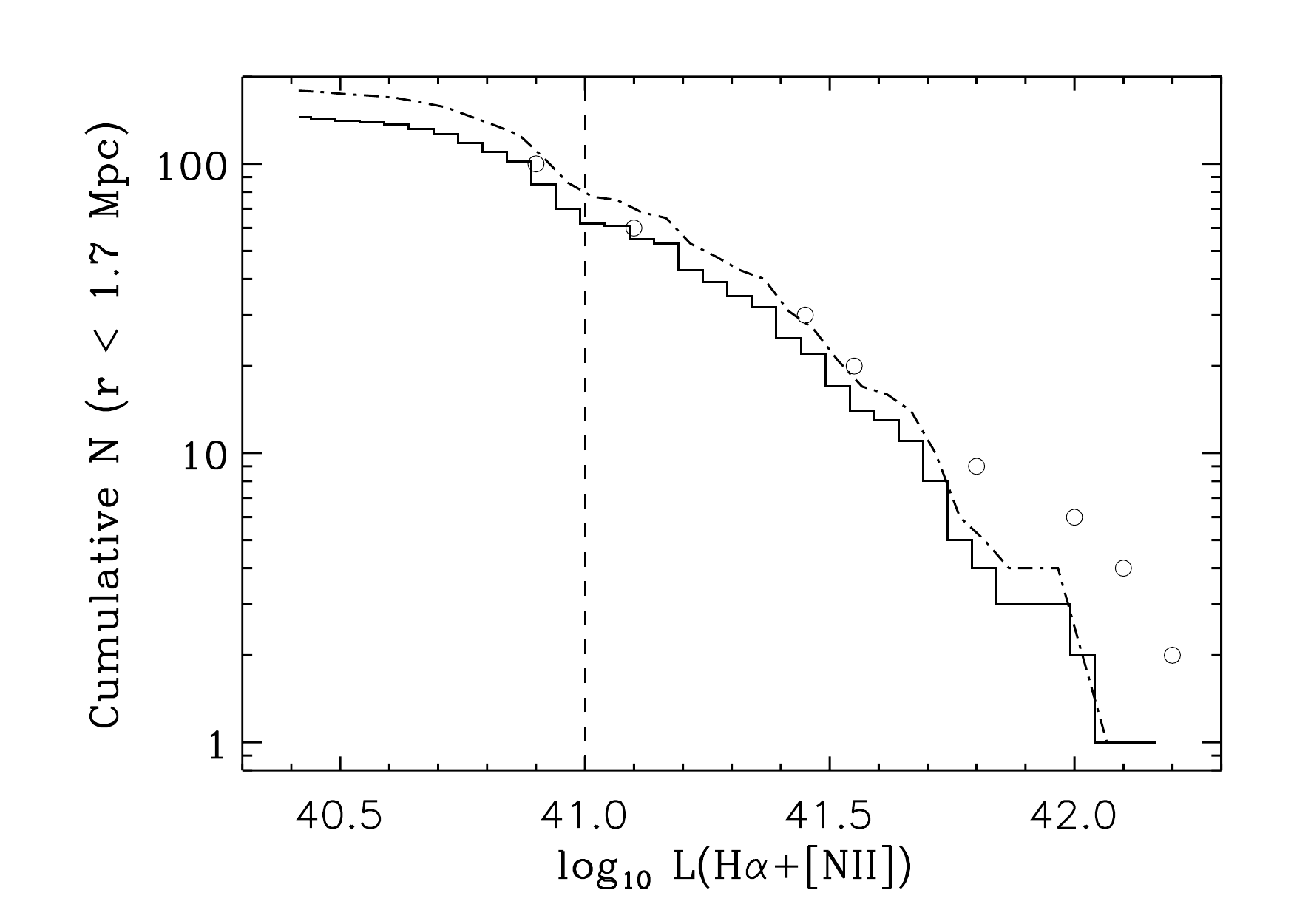}
      \caption{Cumulative luminosity function (LF) within a radius of 0.8\,Mpc (left) and r$_{vir}$\,=\,1.7 Mpc (right). The solid line corresponds to our galaxy counts; the big open circles are sparsely sampled points from the \cite{Kodama2004} LF. The dotted-dashed line in the right plot corresponds to our LF corrected for the incomplete area coverage  within r$_{vir}$. The dashed line corresponds to the completeness limit of our data, $\log$ L(\Ha$+$\nii)\,$\simeq$\,41.
              }
         \label{fig:lum_func}
   \end{figure*}

In order to verify the accuracy and performance of our photometry, we have built the \Ha\,$+$\,\nii\ luminosity functions computed from sources within a central area of 0.8\,Mpc and within r$_{vir}$\,=\,1.7\,Mpc. These are depicted in Fig.~\ref{fig:lum_func}. These cumulative functions  can be compared with those of \cite{Kodama2004} (hereafter K04). However, while the comparison is direct at the smaller radius, at the larger one we  need to take into account the fact that the coverage of our observations within r$_{vir}$ is incomplete towards East and South--East, around 81\% of the full circle of r$_{vir}$. We have performed a simple area correction to account for this, shown as a dotted-dashed line in the right panel of Fig.~\ref{fig:lum_func}. The completeness of our sample, as measured by the peak of the luminosity histogram, is \mbox{$\log$\,L(\Ha$+$\nii)\,$\simeq$\,41}, very similar to that of K04.

An evident discrepancy between our data set and that of K04 is found at the high-luminosity end, \mbox{$\log$\,L(\Ha$+$\nii)\,$\gtrsim$\,41.8}. This is observed, not only in the larger area but also in the smaller one, and hence it cannot be attributed to our incomplete area coverage. However, it should be taken into account that K04 results are based on different technique, namely a standard narrow-band filter (the Subaru/Suprime-Cam NB$_{912}$ filter with $\lambda_{eff}$\,$=$\,9139\,\AA\ and FWHM\,$=$\,134\,\AA) plus broad-band $BRz'$ filters and hence discrepancies in individual objects can be expected. Of course this affects only a few objects (just 4 objects down to \mbox{$\log$\,L(\Ha$+$\nii)\,$\simeq$\,41.8}). At lower luminosities and even beyond the completeness limit, the agreement between our cumulative number counts and those of K04 is excellent, reinforcing the confidence in the accuracy of our photometry.

\section{Star-forming galaxies and the AGN population}
\label{sect:SF_AGN}

\begin{figure*}[bth]
   \centering
   %\vspace{-7cm}
   %\hspace{-1cm}
   \includegraphics[scale=0.5]{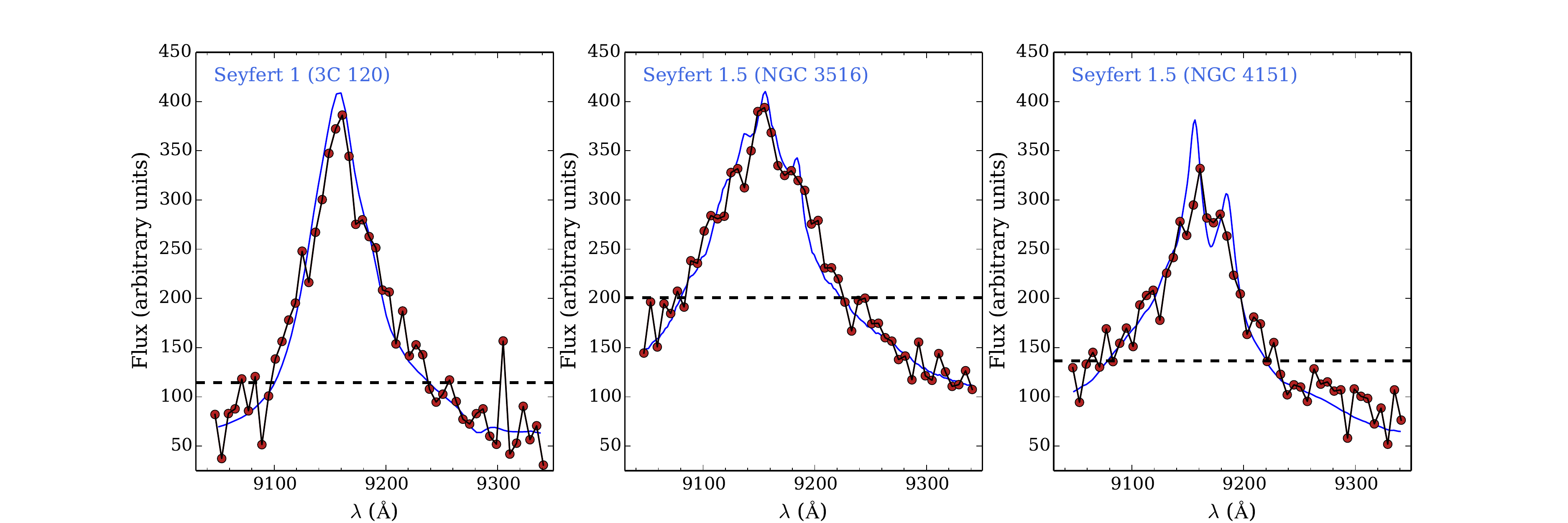}
   \includegraphics[scale=0.43]{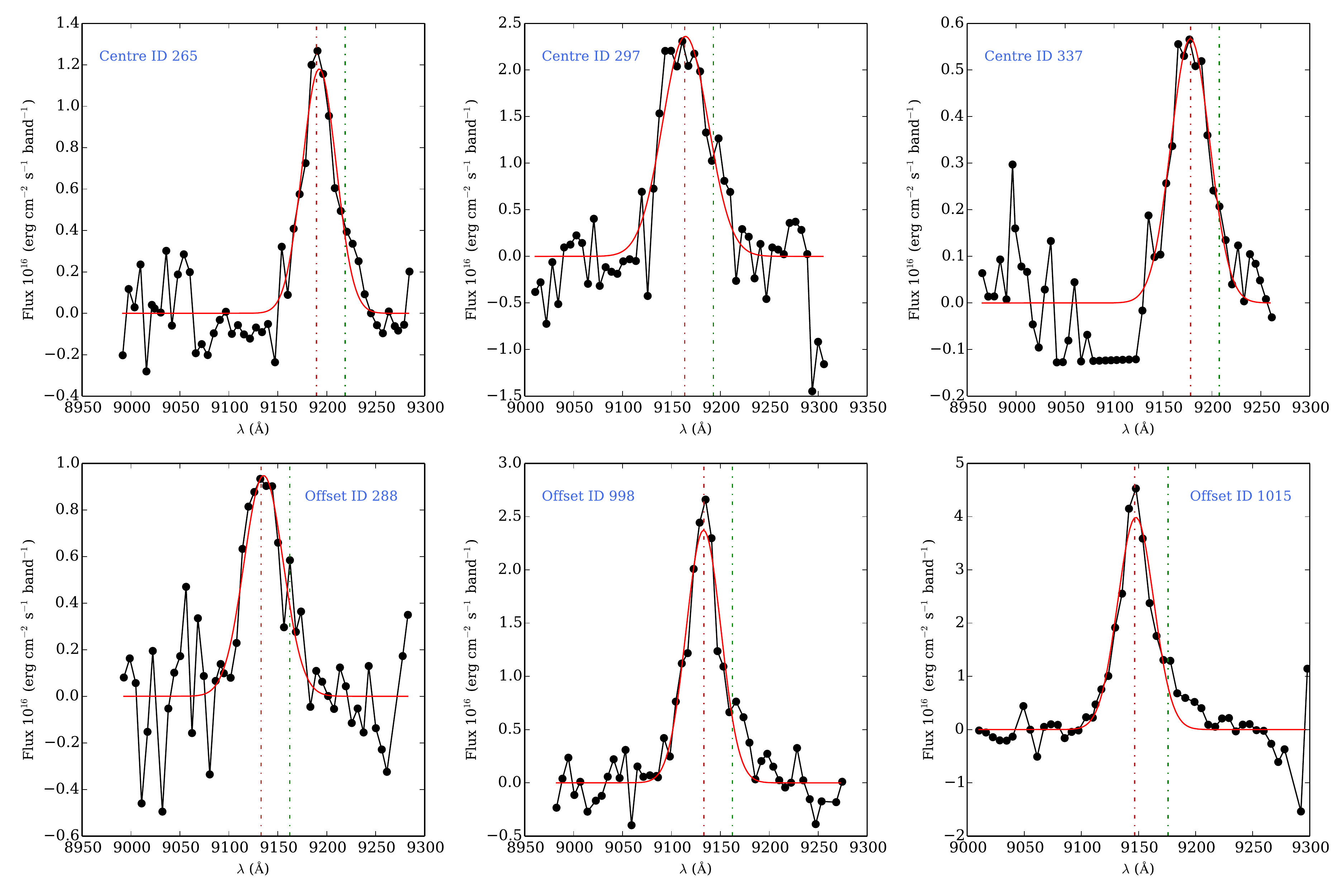}
         \caption{In the top row, real spectra of local universe BLAGNs red-shifted to z\,$=$\,0.395 (blue solid line) and instances of the the simulated pseudo-spectra (red dots and black continuum line) are shown. The dashed horizontal line marks the median value of the pseudo-spectra. See text for details. The middle and bottom rows show actual pseudo-spectra classified as BLAGN. The red and green vertical dash-dotted lines mark the positions of the \Ha\ and \nii\ lines, respectively. The red solid line corresponds to the best fit to a Gaussian profile.}
         \label{fig:blagn}
   \end{figure*}

%\subsection{The AGN population}
We have explored the available mechanisms to separate the population of SF galaxies and AGNs. Broad-line AGNs (BLAGN) show permitted lines with widths of thousands of kilometers per second. By contrast, in narrow-line AGNs (NLAGN), spectra line Doppler widths are  much smaller, typically only a few hundred kilometers per second, which is comparable to or somewhat larger than stellar velocity dispersions. We have verified that the profiles of broad lines are well reproduced in our data by simulating BLAGN pseudo-spectra built from real spectra of local universe Seyfert~1 \citep[3C\,120,][]{Garcia2005} and Seyfert~1.5 galaxies \cite[NGC\,3516;][]{Arribas1997}, \cite[NGC\,4151;][]{Kaspi1996}, that were displaced to the redshift of Cl0024 and convolved with the TF transmission profile (Eq.~\ref{eq:transmission}). Finally, noise components built drawing random values from a normal distribution with zero mean and a standard deviation equal to 10\% of the difference between the peak of the pseudo-spectrum and its median value were  added to the pseudo-spectra signal. In all generated instances of such pseudo-spectra, the broad component of the \Ha\ line was clearly traced as depicted in Fig.~\ref{fig:blagn}.

In order to select candidates to BLAGN we have performed simple Gaussian fits to the pseudo-spectra, setting a low line width threshold of FWHM\,$=$\,36\,\AA\ (6 scan steps) that is approximately 1180\,km s$^{-1}$ at the redshift of Cl0024. The results of the fitting process were carefully inspected, rejecting incorrect or unclear cases (e.g. when two narrow lines were fitted as a single, broad one), arriving to a final list of 25 robust candidates, i.e. 14\% of the sample of ELG. Some cases are depicted in Fig.~\ref{fig:blagn}.

\begin{figure}[htb]
   %\centering
   %\vspace{-7cm}
   \hspace{-0.7cm}
    \includegraphics[width=1.2\columnwidth]{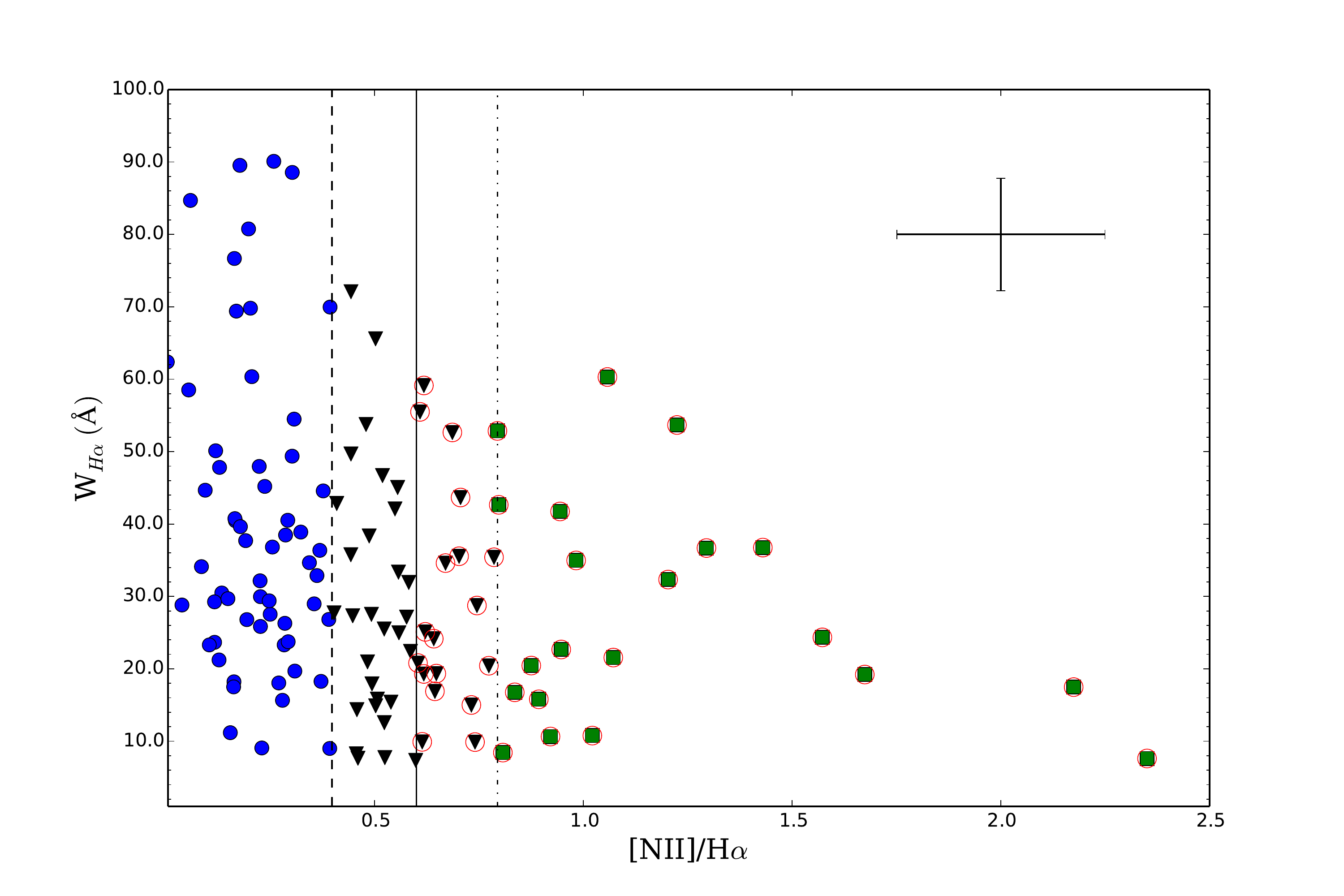}
      \caption{EW$\alpha$n2 diagram showing pure SF galaxies according to S06 criterion (blue dots), pure K01 AGN (green squares) and composite SF+AGN objects (black triangles). Classical AGNs as defined by H97 are denoted by open red circles. The dashed vertical line corresponds to S06 separation criterion, the solid one to the H97 boundary and finally the dashed-dotted line marks the K01 boundary. The error bars in the top right corner correspond to the median errors within our sample of ELG.}
         \label{fig:agn_diag_and_elg_vrad_rclus}
   \end{figure}

In a second step, we have separated NLAGNs from SF galaxies using the ratio between the \nii\ and the \Ha\ lines. Moreover, following \cite{CidFernandes2010} we have combined this ratio with the equivalent width of \Ha\ (W$_{H\alpha}$) in order to break the degeneracy between Seyfert and LINER galaxies, the so-called `EW$\alpha$n2 diagram' as depicted in the left panel of Fig.~\ref{fig:agn_diag_and_elg_vrad_rclus}. The use of W$_{H\alpha}$ is justified by the known fact that Seyfert galaxies tend to have a higher power of the ionising engine with respect to the optical output of the host stellar population. According to \cite{CidFernandes2010}, a sensible Seyfert/LINER boundary can be set at W$_{H\alpha}$\,$=$\,6\,\AA. On the other hand, the fraction of NLAGN depends on the limits set to \niiHa. Different criteria have been proposed: \cite{Ho1997} (hereafter H97) adopted the classical criterion from \cite{Veilleux1987} and consider SF galaxies as those with \niiHa\,$<$\,0.6, and accordingly NLAGNs those in the region \niiHa\,$\geq$\,0.6. This region is populated by objects showing pure AGN spectral features (either Seyferts or LINERS) and `transition' or `composite' objects, interpreted as galaxies whose integrated spectra is the superposition of AGN and SF features. The criterion from \cite{Kewley2001} (K01) is widely used to separate `pure AGNs' ($\log$(\niiHa)\,$\geq$\,-0.10) from SF and composite objects, while those of \cite{Kauffmann2003} (K03) and \cite{Stasinska2006} (S06) are used to separate `pure SF' galaxies from any AGN or composite object. In Tab.~\ref{tab:agn_fractions} the  fractions of NLAGNs found within our sample are shown after applying the different criteria explained above. Taking into account the fraction of BLAGNs found above, the total fraction of AGNs with respect to the ELG population ranges from 26\% to 54\% depending on the diagnostic used.  Unless otherwise stated, the H97 criterion will be used hereafter to separate SF and NLAGN classes. According to this criterion, the fraction of BLAGNs with respect to NLAGNs is 64\%, very similar to the fractions of Seyfert\,1/Seyfert\,2 in the local universe from \citet[][61\%]{Ho1997} and \citet[][60\%]{Sorrentino2006} and the fraction (BLAGN+NLAGN)/ELG $\sim$\,37/\%. 

\begin{table}[h]
\caption{Fraction of NLAGNs in Cl0024 applying different criteria}             % title of Table
\label{tab:agn_fractions}      % is used to refer this table in the text
\centering                          % used for centering table
\begin{tabular}{lccc}        % centered columns (4 columns)
\hline\hline                 % inserts double horizontal lines
Criterion & \niiHa\ & NLAGN  & NLAGN  \\ 
 &  & number  &  Fraction (\%) \\ 
\hline                        % inserts single horizontal line
`pure AGN' (K01) & $\geq$\,0.794 & 21 &  12 \\
`Classical AGN' (H97) & $\geq$\,0.60 & 39 & 22 \\
AGN+composite (K03) & $\geq$\,0.478 & 60 & 34 \\
AGN+composite (S06) & $\geq$\,0.398 & 69 & 40 \\
\hline                                   %inserts single line
\end{tabular}
\end{table}

Regarding the AGN class, all the objects are very likely Seyfert galaxies (W$_{H\alpha}$\,$>$\,6\,\AA). No clear LINER-class objects are detected due to our equivalent width detection limits.

A cautionary note regarding errors in our estimates: due to the large uncertainties quoted for the \nii\ line fluxes, the errors of the \niiHa\ ratios are also generally large; in fact, only 30\% of the sample has relative errors below 50\%, and these objects are in a vast majority AGNs and composite objects. Very large fractional errors in the \niiHa\ ratios tend to be found in objects where the \nii\ line is barely detected, and thus very likely pure SF galaxies. Hence we have eventually decided to keep all sources in the study, regardless of their error in the \niiHa\ ratio.

The fractions  of AGNs with respect to the total number of ELGs obtained in this study are higher than our previous estimates for this cluster \citep[$\sim$\,20\%;][only considering NLAGN]{PerezMartinez2013}, but smaller than those obtained by \cite{Lemaux2010} from \oii\  and \Ha\ measurements in  two clusters at a higher redshift, RXJ1821.6+6827at z\,$\approx$\,0.82 and Cl1604 at z\,$\approx$\,0.9. These authors found that a fraction as large as $\sim$\,68\% of the objects can be classified as AGNs (Seyfert/LINER, using H97 boundaries), and that nearly half of the sample have \oii\ to \Ha\ equivalent width ratios higher than unity, the typical value observed for star-forming galaxies. The fraction of galaxies classified as AGN in their study was reduced to about 33\% for blue galaxies. It is worth mentioning that the results from these authors are derived from a relatively small sample, 19 galaxies (out of a larger sample of 131 \oii\ emitters) compared with ours (174 objects). 

Furthermore, we have compared our result with the fraction of AGNs, computed with respect to the total number of cluster members brighter than a certain magnitude threshold, obtained by \cite{Martini2002} from a deep X-ray \chandra\ observation of the closer, massive cluster A2104 at z\,=\,0.154. They found 6 sources associated with red galaxies within the cluster whose X-ray properties are compatible with being AGN. Notably, only one of them shows optical features of an active nucleus. The authors conclude that at least $\gtrsim$\,5\% of the cluster galaxies with R\,$<$\,20 (the limiting magnitude of their counterparts) harbour an AGN. In order to perform a rough comparison, we have determined the R-band magnitude distribution of our AGN sample using M05 broadband data. The limiting magnitude of the \cite{Martini2002} counterparts corresponds to R\,$=$\,22.3 at the redshift of Cl0024. This value corresponds almost exactly with the peak of the R-band magnitude distribution of our AGNs. Within our sample, 45 AGNs have R\,$\leq$\,22.3.  We have determined the number of galaxies in the M05 catalogue with R\,$\leq$\,22.3 in the redshift interval 0.35\,$<$\,z\,$<$0.45 (both spectroscopic and photometric), within the same area of our AGN sample, obtaining 263 sources. Hence, a crude estimate of the fraction of AGNs within our sample would be 45/263\,$\approx$\,17\%, i.e. more than three times larger than the fractions obtained by means of X-ray estimates. However, once again it must be noticed that the X-ray sample is much smaller (and shallower) than ours, and hence the result should be considered with care. Actually our fraction of AGNs is similar to that obtained by \cite{Lemaux2010}, 20\% at  z\,$\approx$\,0.9. This suggests a lack of redshift evolution of the population of faint AGNs in the range 0.4\,$\lesssim$\,z\,$\lesssim$\,0.9, or at least much milder than that predicted by \cite{Martini2009} for luminous AGNs, $f \sim (1+z)^{5.3}$.

The average of our AGN luminosity distribution (with no extinction correction) is  $\log$L(\Ha)\,$\simeq$\,41.0. At the bright end, we found one AGN with $\log$L(\Ha)\,$\simeq$\,41.7, that corresponds to $\log$L$_X$\,$\sim$\,43 using the L(\Ha) to L$_X$ relation from \cite{Ho2001}. This detection is compatible with the results from \cite{Martini2009} in the 0.3\,$<$\,z\,$<$\,0.6 bin (four AGNs in ten clusters  with $\log$L$_X$\,$\geq$\,43).  

We have cross-matched our ELG catalogue with a list of source detection positions from a \chandra\ ACIS-S observation of 40\,ksec performed in FAINT mode,  kindly provided to us by P.~Tozzi. Five sources (out of 37) have been ascribed to the cluster after cross-matching the X-ray emitter list with the catalogue of spectroscopic redshifts from M05. From these, 3 sources have been detected as ELG in our survey and have been classified as pure AGN or composite according to the \niiHa\ ratio. The remaining 2 sources are detected in our deep TF images but their pseudo-spectra do not show emission line features at a significant level of detection.

%\subsection{Star formation rate}

For the H97 SF galaxies, we have computed the star formation rate (SFR) using the standard assumption of 1 magnitude of extinction at the \Ha\ line \citep[][in a forthcoming paper a more realistic estimation will be done using the Balmer decrement computed with the \Hb\ flux]{Kennicutt1992} and applying the standard luminosity-SFR conversion from \cite{Kennicutt1998}. The histogram of the SFR distribution is depicted in Fig.~\ref{fig:flux_histo} (right panel). The SFR peaks at 0.8\,M$_{\odot}$\,yr$^{-1}$  with a median value of  1.4\,M$_{\odot}$\,yr$^{-1}$. 

The spatial distribution of AGN and SF galaxies in the radial velocity/projected cluster-centric distance space is depicted in  Fig.~\ref{fig:elg_vrad_rclus}. The very central region (r\,$\lesssim$\,250\,Kpc) is almost devoid of ELG, indicating an almost complete quenching of the star formation or AGN activity. AGNs are observed in both cluster components (A and B as defined in sect.~\ref{sect:redshift}), but are more abundant in the main structure. Beyond this point, the analysis of the distribution of galaxies relative to the cluster-centric distance has a limited validity, since on the one hand it does not account for the limited area coverage of our observations at large radii, and on the other it does not provide information of the importance of star formation or AGN activity relative to the density of galaxies in the surrounding environment. A thorough analysis of the dependency of the SFR and specific SFR with the local galaxy density (characterised by the $\Sigma_5$ parameter) will be presented in P\'erez-Mart\'{\i}nez et al. (in prep.).

\begin{figure}[htb]
   %\centering
   \hspace{-0.20cm}
   \includegraphics[scale=0.47]{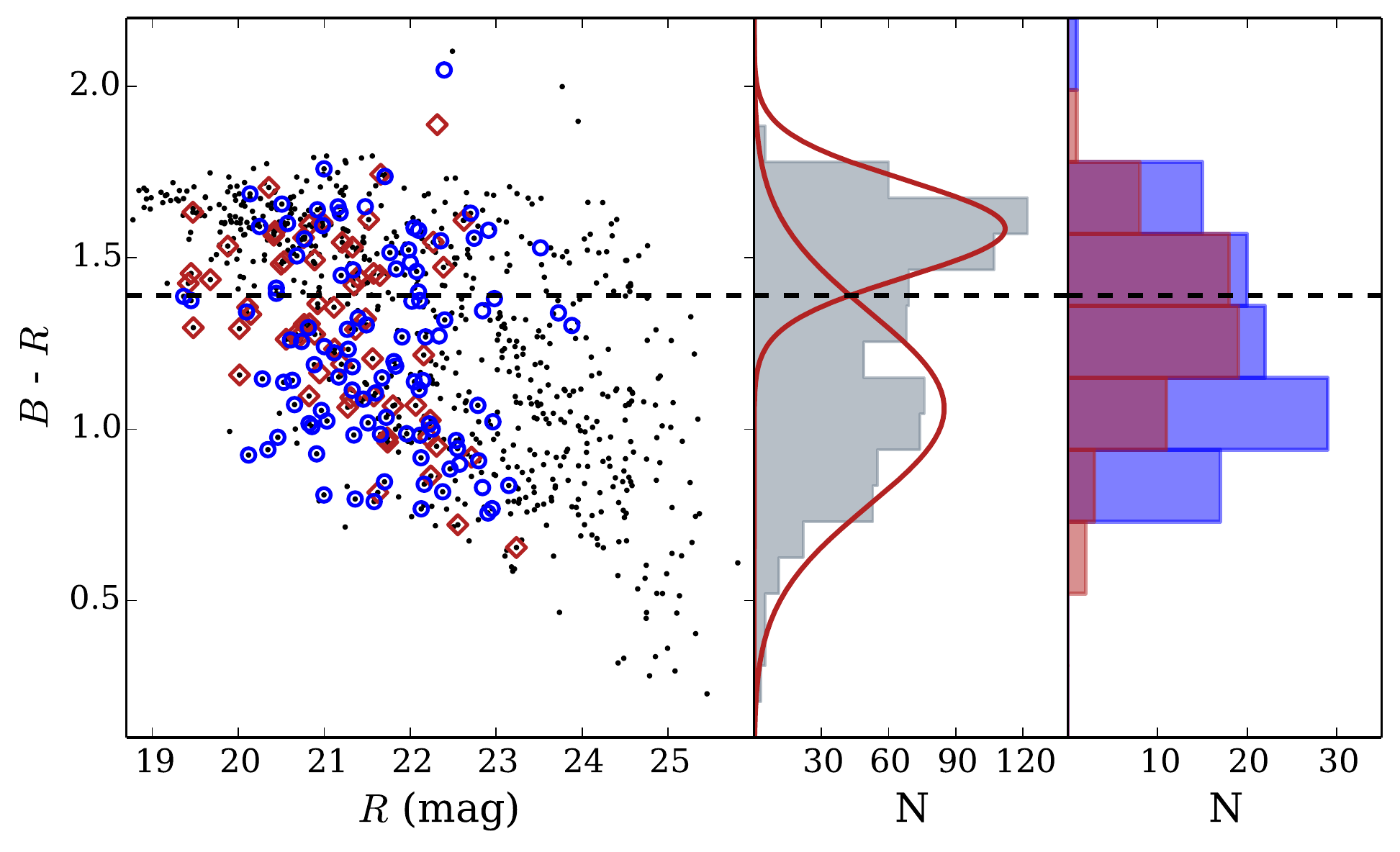}
      %\hspace{-1.5cm}
      \caption{\textit{Left}: $B - R$ vs. $R$ colour--magnitude  diagram. The AGN are represented by red diamond symbols, while blue circles are SF galaxies. The black dots represent the cluster control sample from M05. The dashed line corresponds to the separation between the red sequence and blue cloud populations (see text). \textit{Middle}: histogram of the control sample. The bi-modality of the distribution of optical colours is evident. The red curves are the best Gaussian fits to the distribution \textit{Right}: colour  histograms for AGNs (red) and SF (blue) galaxies. It is likely that both populations are drawn from different distributions. 
              }
         \label{fig:cmr}
   \end{figure}

Finally, we have plotted the colour--magnitude diagram (CMD) for the ELG sample, along with a control sample of cluster galaxies from M05. This control sample, comprising 792 galaxies, has been drawn from the main catalogue by choosing all the objects having valid $BR$ photometry, within the same sky area as our sample and fulfilling 0.35\,$<$\,z\,$<$\,0.45 (either spectroscopic or photometric). The \textit{K}-correction has been applied to the M05 magnitudes by means of \texttt{kcorrect v4\_2} \citep{Blanton2007}. The rest-frame $B - R$ vs. $R$ CMD is depicted in Fig.~\ref{fig:cmr}. The bi-modality in the distribution of optical colours is clearly noticed, with a red sequence well developed. We have defined a boundary between the red sequence and the blue cloud as the intersection of the Gaussian functions resulting from the fit to each colour peak ( $B - R = 1.39$). We have investigated the colour distributions of AGNs and SF galaxies (right panel in Fig.~\ref{fig:cmr}). The differences are not outstanding but noticeable. The Kolmogorov-Smirnov test rejects the null hypothesis that both samples are drawn from the same distribution with a \textit{p}--value of 0.09. The colour distribution of SF galaxies reaches a clear maximum in the blue cloud, with a tail towards the red sequence region (either due to dust absorption or to the presence of old, red stellar populations), while the AGN colour distribution tends to peak at the boundary between the two regions, the so-called `green valley', as expected for the population of galaxies hosting active nuclei \citep[see for instance][ and references therein]{Povic2012}.

\section{Cluster dynamics from emission line galaxies }
\label{sect:cluster_dynamics}

As a final test of the reliability and usefulness of our data beyond the main drivers of the GLACE survey, we have applied the caustic technique \citep{Diaferio1997,Diaferio1999} using the \texttt{CausticApp} code (Serra \& Diaferio, in prep.) to explore the possibility of using ELG radial velocity data to trace the cluster mass. This method, unlike the traditional Jeans approach, does not rely on the assumption of dynamical equilibrium, and can be used to estimate the cluster mass even in non-equilibrium regions. Thus, in principle, it is well suited to be applied to ELGs, that are likely not in dynamical equilibrium. 

To briefly summarise the physical idea behind the caustic technique, we will follow \cite{Serra2011}: in hierarchical clustering models of structure formation, clusters form by the aggregation of smaller systems. This accretion does not take place purely radially, but particles within the falling clumps have velocities with a substantial non-radial component.
The rms velocity $\langle v^2 \rangle$ is due to the gravitational potential of the cluster and the groups where the galaxies reside, and to the tidal fields of the surrounding region. 

In a redshift diagram (i.e. line-of-sight velocity v$_{los}$ vs. projected distance to the center R$_p$), cluster galaxies populate a region with a characteristic trumpet shape with decreasing amplitude $\mathcal{A}({\rm R}_p)$ with increasing R$_p$. This amplitude is related to $\langle {\rm v}^2 \rangle$. \cite{Diaferio1997} identified this amplitude with the average component along the line of sight of the escape velocity at the three-dimensional radius r\,$=$\,R$_p$. The projection along the line of sight involves a function which depends on the anisotropy parameter $\beta({\rm r}) = 1 - (\langle {\rm v}^2_{\theta} \rangle + \langle {\rm v}^2_{\phi}\rangle)/2\langle {\rm v}^2_{\rm r} \rangle$. Here, $\langle {\rm v}^2_{\theta} \rangle$, $\langle {\rm v}^2_{\phi}\rangle$ and $\langle {\rm v}^2_{\rm r}\rangle$ are the longitudinal, azimuthal and radial components of the velocity \textbf{v} of a galaxy, respectively and the brackets represent the average over a local volume at a given position \textbf{r}. Considering the cluster rotation negligible, $\langle {\rm v}^2_{\theta} \rangle = \langle {\rm v}^2_{\phi}\rangle = \langle {\rm v}^2_{los}\rangle$, where $\langle {\rm v}^2_{los}\rangle$ is the line-of-sight component of the velocity. 

Following the results of \cite{Diaferio1997}, the square of the caustic amplitude $\mathcal{A}^2({\rm r}) = \langle {\rm v}^2_{esc,los} \rangle $ and relating the escape velocity to the cluster gravitational potential as  $\langle {\rm v}^2_{esc} \rangle = -2\phi({\rm r})$, it can be shown that

\begin{equation}
GM(<{\rm r}) = \int_0^{\rm r} \mathcal{A}^2({\rm r})g(\beta)\mathcal{F}({\rm r})\,{\rm dr}
\label{eq:caustic_integ}
\end{equation}

\noindent where $\mathcal{F}({\rm r}) = -2 \pi G \rho ({\rm r}){\rm r}^2/ \phi ({\rm r})$, with $\rho ({\rm r})$ the mass density profile and $g(\beta) = (3 - 2\beta ({\rm r}))/(1 - \beta ({\rm r}))$. In order to solve Eq.~\ref{eq:caustic_integ}, it is assumed that the product $\mathcal{F}_{\beta}({\rm r})$\,$=$\,$\mathcal{F}({\rm r}) g(\beta)$ varies slowly with r, and hence it is possible to replace it by a constant parameter $\mathcal{F}_{\beta}$. It is easy to show that $\mathcal{F}({\rm r})$ is a slowly varying function of r in hierarchical clustering scenarios. The assumption that  $\mathcal{F}_{\beta}({\rm r})$ is also a slow function of r, is somewhat stronger and is demonstrated in \cite{Serra2011}. 

Therefore, the mass profile could be estimated as:

\begin{equation}
GM(<{\rm r}) = \mathcal{F}_{\beta} \int_0^{\rm r} \mathcal{A}^2( {\rm r})\,{\rm dr}. 
\label{eq:caustic_integ_approx}
\end{equation}

The \texttt{CausticApp} is a graphical interface that allows an easy use of the caustic technique. The technique uses a binary tree according to a hierarchical method to arrange the galaxies in a catalogue according to their projected binding energy. By cutting the tree at the appropriate thresholds \citep[details in][]{Serra2011}, it provides a set of candidate members, the cluster centre, velocity dispersion and size. With that information a redshift diagram is created and the caustics located. Finally, the caustic mass profile is estimated.

\begin{figure*}[htb]
   %\centering
   %\vspace{-7cm}
   \hspace{-0.5cm}
   \includegraphics[scale=0.4]{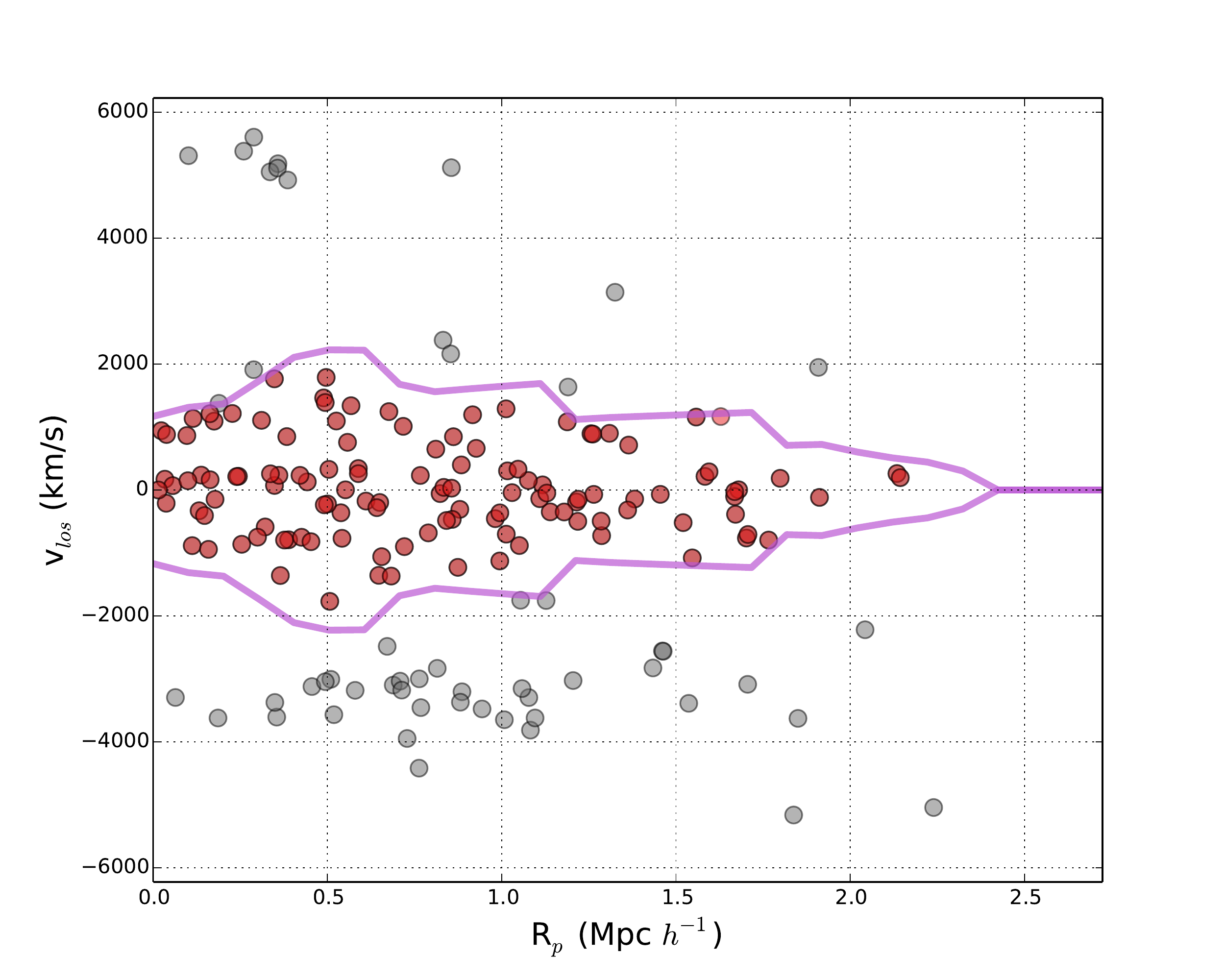}
   \hspace{-1.cm}
   \includegraphics[scale=0.4]{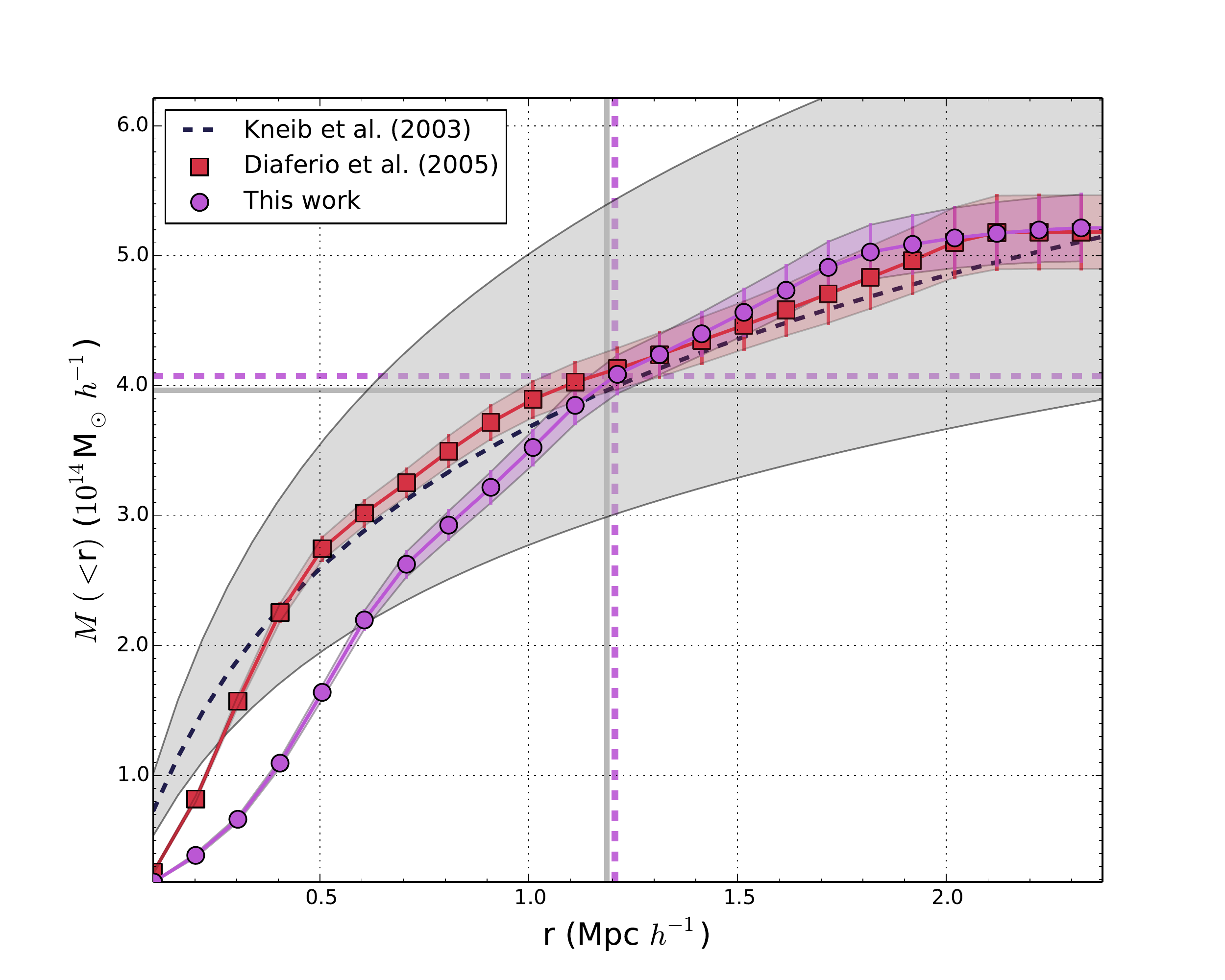}
      \caption{\textit{Left}: redshift diagram showing the selected cluster members (red dots, coincident with those in the main structure identified in sect.~\ref{sect:redshift}) and the caustics (purple solid lines). \textit{Right}: caustic-ELG mass profile (purple circles) computed by integrating the squared caustic amplitude profile; NFW-best-fit mass profile to the weak-lensing mass from \cite{Kneib2003} (black dashed line); caustic analysis of the catalog from \cite{Diaferio2005} (red squares). The vertical and horizontal lines are M$_{200}$ and r$_{200}$ of the caustic-ELG (purple) and weak-lensing (grey) mass profiles.
              }
         \label{fig:caustics}
   \end{figure*}

We have run the code with our complete data set of 174 ELGs, setting a conservative filling factor $\mathcal{F}_{\beta}=0.5$. The center of the cluster computed by the code is $\alpha_{\rm ELG}=0^h\,26^m\,40.8^s$, $\delta_{\rm ELG}=17^\circ\,9'\,21.6''$ and the final members are 121. The redshift diagram with the location of the caustics and the cluster mass profile are depicted in the left and right panels of Fig.~\ref{fig:caustics}, respectively. 

Fig.~\ref{fig:caustics} also shows other two mass profiles estimated from ZwCl 0024.0+1652 data: a weak-lensing mass profile from \cite{Kneib2003} and another caustic mass profile, obtained with the larger galaxy catalog with 333 galaxies within the redshift range $z$=[0.35,0.45], used by \cite{Diaferio2005}. For this analysis we kept the cluster center used in \cite{Diaferio2005}: $\alpha_{\rm D05}=0^h\,26^m\,45.9^s$, $\delta_{\rm D05}=17^\circ\,9'\,41.1''$. The number of members identified in this case is 251. Despite the number difference, our sample is able to yield a very good estimate of the center. Indeed, the offset between ($\alpha_{\rm ELG}, \delta_{\rm ELG}$) and ($\alpha_{\rm D05}, \delta_{\rm D05}$) is negligible.

The agreement among all of the three mass profiles is remarkable at r=[1,2.3] Mpc\,$h^{-1}$. In the inner regions, the ELG-caustic mass profile is considerably lower than the other two, due to lack of galaxies in the very central regions of the cluster. In most clusters ELGs avoid the cluster center, which translates into a caustic amplitude that tends to underestimate the mass profile in the inner regions. As the mass profile is cumulative, this underestimate propagates into the final mass estimates. However, the cumulative caustic mass profile from ELGs in Fig \ref{fig:caustics} shows that the central mass does not contribute substantially to the total cluster mass at large radii. This suggests that in this cluster the velocity field is well sampled by the ELGs in the outskirts; consequently, the ELG-caustic mass profile is consistent with the estimates from previous studies.

With the caustic method, we derive r$_{200}$\,$=$\,(1.21 $\pm$ 0.1)\,Mpc $h^{-1}$ and M$_{200}$\,$=$\,(4.1\,$\pm$\,0.2)\,$\times$\,10$^{14}$\,M$_{\odot}\, h^{-1}$, where the uncertainties are due to the uncertainties in the mass profile. For comparison, \cite{Kneib2003}, from weak-lensing analysis, derived r$_{200}$\,$=$\,(1.19\,$\pm$\,0.9)\,Mpc\,$h^{-1}$ and M$_{200}$\,$=$\,(3.97\,$^{+0.8}_{-0.7}$)\,$\times$\,10$^{14}$\,M$_{\odot}\,h^{-1}$. Therefore, there is very good agreement between our results and those obtained by means of a radically different approach.

\section{Summary and Conclusions}

Aimed at exploring the technical feasibility of the GLACE project, we have carried out a TF survey of the intermediate redshift cluster ZwCl 0024.0+1652 (Cl0024 hereafter) at z\,=\,0.395, targetting the \Hanii\ line complex.  We have performed two pointings, covering approximately 2 virial radii and a velocity field of approximately $\pm$\,4000\,km s$^{-1}$. Cl0024 is a very well studied cluster, and hence a good test bench to explore the capabilities, performance and limitations of the TF tomography method. In this paper we present the main technical aspects related to the creation of the catalogue of \Ha\ emitters, along with some simulations aimed at confirming the validity  (and limitations) of our methods. We present a set of scientific results that can be used as a benchmark of the performance of our techniques upon comparison with previous works. These results include the spatial and redshift distribution of cluster galaxies, the \Ha\ luminosity function, the population of star-forming (SF) galaxies and AGNs and finally, an attempt to make estimations on the mass and velocity dispersion of the cluster based on our emission-line data. The main results are:

\begin{itemize}
\item A catalogue of 174 unique cluster emitters has been compiled. The completeness limit of our \Ha\ line sample is $f_{H\alpha} \sim 0.9 \times 10^{-16}$\,erg s$^{-1}$ cm$^{-2}$, a flux consistent or below the GLACE requirement.

\item We have compared our redshift estimates for our galaxies, as given by the \Ha\ line position in the pseudo-spectra, with spectroscopic redshifts from \cite{Moran2005}. There is an excellent agreement: the redshift error defined as $\vert z_{TF} - z_{spec} \vert / ( 1 + z_{spec}) $ is on average 0.002 (median value 0.0005) with a maximum value of 0.02. Hence, we can consider the TF-derived redshifts as of spectroscopic quality. 

\item The  redshift distribution of the cluster galaxies allows us to identify two interacting structures already identified in spectroscopic surveys \citep{Czoske2002}. Moreover, we suggest the existence of a third interacting group (previously not identified as such, but as part of the surrounding field galaxy population), although the clustering of the sources can be due to an instrumental effect (uneven spatial coverage at the shortest and longest wavelengths). 

\item The \Ha\ luminosity function is well aligned with previous estimates from \cite{Kodama2004} up to the completeness limit of the sample, $\log$\,L(\Ha$+$\nii)\,(erg s$^{-1}$)\,$\simeq$\,41. 

\item  Broad-line AGNs (BLAGN) can be separated from narrow-line ones (NLAGN) or SF galaxies. In order to select candidates to BLAGN we have performed simple Gaussian fits to the pseudo-spectra, arriving to a final list of 25 robust candidates, i.e. 14\% of the sample of ELGs. Furthermore, our observing technique allowed us to deblend the \Ha\ and \nii\ lines. In fact, in an outstanding fraction of the objects, both lines appear clearly separated even in a visual inspection of the pseudo-spectra. We have used this deblending capability  to discriminate the population of NLAGNs from the SF galaxies. The fraction of NLAGNs depends on the criterion used. If a `classical' separation criterion is applied \citep[][H97]{Ho1997}, the fraction of NLAGNs is $\sim$\,22\%. The total fraction of AGN (BLAGN+NLAGN) is  $\sim$\,37\% or around 17\% of the cluster population brighter than R\,$=$\,22.3, similar to estimates derived \oii\ observations z\,$\approx$\,0.9. 
This suggests a lack of redshift evolution of the population of faint AGNs in the range 0.4\,$\lesssim$\,z\,$\lesssim$\,0.9, or at least much milder than previous predictions.

\item The star formation rate (SFR), computed using the standard assumption of 1 magnitude of extinction at the \Ha\ line and the standard luminosity-SFR conversion from \cite{Kennicutt1998}, peaks at 0.8\,M$_{\odot}$\,yr$^{-1}$  with a median value of  1.4\,M$_{\odot}$\,yr$^{-1}$.

\item We have found that the innermost region of the cluster (r\,$\lesssim$\,250\,Kpc) is virtually devoid of ELGs, indicating an almost complete quenching of the star formation and AGN activity, as expected for a cluster at this redshift.

\item We have used the dynamical information (radial velocities) provided by our ELGs to estimate the mass, radius and velocity dispersion of Cl0024 using the technique of caustics in the redshift space \cite{Diaferio1997,Diaferio1999} as implemented in the \texttt{CausticApp} application from \cite{Serra2011}. With the caustic method, we derive r$_{200}$\,$=$\,(1.21 $\pm$ 0.1)\,Mpc $h^{-1}$ and M$_{200}$\,$=$\,(4.1\,$\pm$\,0.2)\,$\times$\,10$^{14}$\,M$_{\odot}\, h^{-1}$, in very good agreement with the weak-lensing estimate from \cite{Kneib2003}.

\end{itemize}

The results outlined above indicate that the TF tomography technique is a powerful tool for the proposed research and that the devised observing methods and reduction and analysis procedures yield results that meet or exceed the GLACE requirements. The tunable filters of the OSIRIS instrument at the 10.4m GTC telescope constitute a powerful tool to explore galaxy cluster, allowing to derive important properties of the members, and even to investigate the cluster dynamics.

%\section{ToDo's}
%
%Check SSFR vas stellar mass. For early type objects, these can be either in the main sequence (indicating that quenching has not started yet) or below it, indicating the onset of SF quenching.

\bibliographystyle{aa} % style aa.bst 
\bibliography{referencias} % your references Yourfile.bib

%\appendix

%\section{Catalogue of emitters of Cl0024}

\longtab{5}{
\begin{center}
\begin{longtable}{l c c c c c c c c p{1cm}}

\caption{\label{tab:cldata} Catalogue of unique ELGs of Cl0024. The identifiers with 'C' correspond to sources gathered from the central pointing, while those with 'O' correspond to sources gathered from the offset one.} \\
\hline \hline
\multirow{2}{*}{ID} & RA J2000 & Dec J2000 & \multirow{2}{*}{z} & f$_{H\alpha}$\,10$^{-17}$ & f$_{[NII]}$\,10$^{-17}$ & EW$_{H\alpha}$ & EW$_{H\alpha+[NII]}$ & \multirow{2}{*}{Type} \\
 & (00:26:ss.ss) & (+17:mm:ss.s) & &  (erg\,cm$^{-2}$s$^{-1}$) & (erg\,cm$^{-2}$s$^{-1}$) & (\AA) & (\AA) \\
\hline
\endfirsthead
\caption{continued.} \\

\hline \hline
\multirow{2}{*}{ID} & RA J2000 & Dec J2000 & \multirow{2}{*}{z} & f$_{H\alpha}$\,10$^{-17}$ & f$_{[NII]}$\,10$^{-17}$ & EW$_{H\alpha}$ & EW$_{H\alpha+[NII]}$ & \multirow{2}{*}{Type} \\
 & (00:26:ss.ss) & (+17:mm:ss.s) & & (erg\,cm$^{-2}$s$^{-1}$) & (erg\,cm$^{-2}$s$^{-1}$) & (\AA) & (\AA) \\
\hline
\endhead
\hline
\endfoot 
   4\_C    &   34.27 &  06:04.5  & 0.3820 & 33$\pm$6   & 2$\pm$2     & 84.7 & 88.9& SF \\		
  12\_C   &   42.98 &  06:12.2  & 0.3999 & 18$\pm$6   & -	    & 38.9 & - & BLAGN \\		
  24\_C   &   28.93 &  06:20.8  & 0.3982 & 7$\pm$3    & 5$\pm$3     & 9.9 & 15.6& NLAGN \\		
  35\_C   &   23.84 &  06:26.9  & 0.3955 & 10$\pm$6   & 4$\pm$4     &  9.0  &  12.3& SF \\		
  60\_C   &   38.72 &  06:46.9  & 0.3967 & 12$\pm$3   & -	    &  42.3  &  - & SF \\  	
  70\_C   &   29.05 &  06:59.1  & 0.3785 & 10$\pm$4   & 5$\pm$7     &  17.9  &  26.4& SF \\  	
  72\_C   &   19.39 &  07:03.0  & 0.3993 & 35$\pm$11  & 18$\pm$10    &  17.4  &  26.6& BLAGN \\ 	
  73\_C   &   42.31 &  07:04.7  & 0.3949 & 7$\pm$2    & 8$\pm$2     &  36.7  &  87.2& NLAGN \\  	
  87\_C   &   14.95 &  07:14.5  & 0.3803 & 13$\pm$5   & 14$\pm$8    &  21.6  &  43.6& NLAGN \\  	
  96\_C   &   34.49 &  07:15.4  & 0.3832 & 37$\pm$11  & -	    &  35.5  &  - & SF \\  	
  97\_C   &   34.88 &  07:15.4  & 0.4014 & 10$\pm$3   & 5$\pm$3     &  14.9  &  22.3& SF \\  	
  105\_C  &   14.79 &  07:19.8  & 0.3832 & 60$\pm$14  & -     &  62.4  &  - & SF \\  	
  106\_C  &   30.99 &  07:22.0  & 0.3807 & 25$\pm$7   & 6$\pm$4     &  30.0  &  36.5& SF \\  	
  138\_C  &   34.22 &  07:42.8  & 0.3812 & 33$\pm$8   & 12$\pm$6    &  36.4  &  49.3& SF \\  	
  143\_C  &   39.26 &  07:46.5  & 0.3784 & 14$\pm$4   & 1$\pm$3     &  34.1  &  36.9& SF \\  	
  147\_C  &   41.17 &  07:47.0  & 0.3888 & 13$\pm$3   & 2$\pm$2     &  40.5  &  46.5& SF \\  	
  153\_C  &   21.48 &  07:55.0  & 0.3774 & 16$\pm$5   & 11$\pm$5    &  43.7  &  75.2& NLAGN \\  	
  163\_C  &   44.93 &  08:10.1  & 0.3955 & 25$\pm$18  & 22$\pm$19    &  6.7  &  12.7& BLAGN \\  	
  193\_C  &   40.10 &  08:21.8  & 0.4008 & 6$\pm$2    & 4$\pm$5     &  15.0  &  26.0& NLAGN \\  	
  196\_C  &   27.09 &  08:23.8  & 0.3812 & 15$\pm$6   & 13$\pm$9    &  15.8  &  30.0& NLAGN \\  	
  200\_C  &   42.90 &  08:23.8  & 0.3962 & 11$\pm$3   & 8$\pm$5     &  35.6  &  58.8& NLAGN \\  	
  219\_C  &   29.81 &  08:36.4  & 0.3993 & 59$\pm$15  & 21$\pm$10    &  62.1  &  81.3& BLAGN \\ 	
  224\_C  &   16.48 &  08:40.3  & 0.3943 & 7$\pm$1    & 2$\pm$2     &  44.6  &  59.5& SF \\  	
  227\_C  &   35.73 &  08:40.6  & 0.4003 & 13$\pm$3   & 16$\pm$7    &  32.3  &  70.4& NLAGN \\  	
  246\_C  &   23.79 &  08:49.9  & 0.4002 & 17$\pm$6   & 5$\pm$9     &  15.7  &  20.0& SF \\  	
  253\_C  &   23.79 &  08:50.2  & 0.3810 & 19$\pm$4   & 6$\pm$3     &  54.5  &  69.7& SF \\  	
  255\_C  &   38.29 &  08:51.4  & 0.4007 & 7$\pm$2    & 3$\pm$1     &  49.7  &  70.0& SF \\  	
  263\_C  &   27.70 &  08:51.5  & 0.3991 & 48$\pm$9   & 30$\pm$11    &  59.1  &  90.7& NLAGN \\ 	
  265\_C  &   33.82 &  08:56.0  & 0.4002 & 16$\pm$3   & 8$\pm$3     &  47.7  &  67.8& BLAGN \\  	
  280\_C  &   33.83 &  08:56.6  & 0.3910 & 11$\pm$3   & 11$\pm$6    &  22.7  &  44.0& NLAGN \\  	
  285\_C  &   18.64 &  08:57.1  & 0.3935 & 101$\pm$14 & 5$\pm$3     &  425.6  &  437.2& SF \\	
  287\_C  &   18.63 &  08:57.1  & 0.3942 & 42$\pm$15  & 12$\pm$8    &  23.3  &  29.9& SF \\  	
  290\_C  &   42.37 &  09:02.2  & 0.3798 & 6$\pm$3    & 4$\pm$2     &  16.9  &  28.0& SF \\  	
  293\_C  &   30.79 &  09:02.5  & 0.4008 & 4$\pm$1    & 7$\pm$3     &  24.3  &  62.9& NLAGN \\  	
  297\_C  &   11.22 &  09:02.7  & 0.3962 & 31$\pm$12  & 32$\pm$21    &  12.1  &  24.5& NLAGN \\ 	
  303\_C  &   30.80 &  09:02.9  & 0.3870 & 5$\pm$2    & -	    &  28.3  &  - & BLAGN \\  	
  308\_C  &   37.99 &  09:04.4  & 0.3995 & 40$\pm$7   & 8$\pm$4     &  69.8  &  82.3& SF \\  	
  314\_C  &   41.09 &  09:05.3  & 0.3959 & 6$\pm$2    & 1$\pm$2     &  30.5  &  34.5& SF \\  	
  317\_C  &   19.72 &  09:10.2  & 0.4009 & 10$\pm$3   & 6$\pm$2     &  22.4  &  35.6& SF \\  	
  320\_C  &   40.69 &  09:11.9  & 0.4002 & 14$\pm$3   & -	    &  89.2  &  - & SF \\  	
  322\_C  &   43.27 &  09:12.6  & 0.4034 & 9$\pm$3    & -	    &  35.5  &  - & SF \\  	
  323\_C  &   43.79 &  09:14.3  & 0.3959 & 8$\pm$2    & -	    &  64.4  &  - & SF \\  	
  332\_C  &   34.29 &  09:14.5  & 0.3811 & 23$\pm$5   & 9$\pm$5     &  70.0  &  92.1& SF \\  	
  337\_C  &   30.97 &  09:14.7  & 0.3985 & 7$\pm$2    & 5$\pm$2     &  81.3  &  128.9& BLAGN \\ 	
  338\_C  &   40.31 &  09:15.7  & 0.3915 & 28$\pm$9   & 14$\pm$10    &  15.8  &  23.9& SF \\ 	
  339\_C  &   31.64 &  09:16.0  & 0.3902 & 13$\pm$4   & 4$\pm$2     &  38.5  &  48.8& SF \\  	
  341\_C  &   28.14 &  09:16.1  & 0.3937 & 27$\pm$8   & 21$\pm$16    &  20.4  &  36.7& NLAGN \\ 	
  343\_C  &   15.28 &  09:17.8  & 0.3992 & 30$\pm$6   & 18$\pm$5    &  55.5  &  90.5& NLAGN \\  	
  345\_C  &   18.26 &  09:17.9  & 0.3955 & 18$\pm$5   & 10$\pm$4    &  25.5  &  38.4& SF \\  	
  353\_C  &   18.25 &  09:17.9  & 0.3951 & 10$\pm$2   & 5$\pm$2     &  38.4  &  55.6& SF \\  	
  358\_C  &   15.25 &  09:18.2  & 0.4193 & 8$\pm$3    & -	    &  26.5  &  - & SF \\  	
  359\_C  &   41.15 &  09:18.6  & 0.4202 & 16$\pm$5   & -     &  48.9  &  - & SF \\  	
  364\_C  &   28.52 &  09:20.5  & 0.4212 & 7$\pm$3    & -     &  35.7  &  - & SF \\  	
  366\_C  &   28.52 &  09:20.9  & 0.3992 & 67$\pm$12  & 17$\pm$8    &  90.1  &  109.3& SF \\ 	
  374\_C  &   31.31 &  09:22.4  & 0.4187 & 6$\pm$3    & -	    &  12.8  &  - & SF \\  	
  382\_C  &   33.50 &  09:23.7  & 0.4040 & 8$\pm$2    & 4$\pm$3     &  25.0  &  38.3& SF \\  	
  388\_C  &   40.08 &  09:24.1  & 0.3783 & 11$\pm$3   & 1$\pm$2     &  58.5  &  61.5& SF \\  	
  405\_C  &   32.32 &  09:24.2  & 0.3936 & 114$\pm$30 & 66$\pm$24    &  27.1  &  42.1& SF \\ 	
  408\_C  &   33.51 &  09:24.3  & 0.3943 & 70$\pm$25  & 34$\pm$14    &  21.0  &  31.2& NLAGN \\ 	
  409\_C  &   35.93 &  09:24.5  & 0.3958 & 26$\pm$16  & 21$\pm$12    &  8.4  &  15.2& SF \\  	
  410\_C  &   40.99 &  09:24.6  & 0.3930 & 11$\pm$4   & 2$\pm$5     &  18.2  &  21.0& SF \\  	
  422\_C  &   34.09 &  09:26.0  & 0.3933 & 67$\pm$14  & 37$\pm$12    &  45.1  &  68.1& SF \\ 	
  423\_C  &   35.39 &  09:26.3  & 0.4199 & 14$\pm$5   & -    &  38.3  &  - & SF \\  	
  424\_C  &   24.34 &  09:26.6  & 0.3908 & 51$\pm$10  & 21$\pm$9    &  42.8  &  59.1& SF \\  	
  433\_C  &   24.34 &  09:26.9  & 0.4004 & 13$\pm$3   & 2$\pm$2     &  47.8  &  53.2& SF \\  	
  437\_C  &   39.84 &  09:28.7  & 0.4189 & 23$\pm$10  & -	    &  44.6  &  - & SF \\  	
  443\_C  &   42.44 &  09:32.2  & 0.3961 & 12$\pm$3   & 4$\pm$5     &  29.0  &  39.7& SF \\  	
  450\_C  &   34.55 &  09:32.8  & 0.3942 & 12$\pm$3   & 3$\pm$2     &  45.2  &  55.0& SF \\  	
  451\_C  &   35.48 &  09:36.0  & 0.3911 & 25$\pm$7   & 5$\pm$4     &  26.8  &  31.8& SF \\  	
  452\_C  &   44.08 &  09:37.9  & 0.3945 & 21$\pm$8   & 8$\pm$10     &  18.3  &  24.8& SF \\ 	
  453\_C  &   39.85 &  09:45.5  & 0.4016 & 6$\pm$3    & 3$\pm$3     &  7.7  &  11.2& SF \\		
  456\_C  &   24.84 &  09:47.2  & 0.3966 & 52$\pm$13  & 48$\pm$20    &  31.9  &  58.5& BLAGN \\ 	
  559\_C  &   40.91 &  09:48.5  & 0.3870 & 7$\pm$3    & 9$\pm$6     &  10.1  &  23.0& BLAGN \\  	
  560\_C  &   18.15 &  09:48.7  & 0.3945 & 29$\pm$6   & 13$\pm$5    &  72.1  &  99.9& SF \\  	
  571\_C  &   18.15 &  09:48.7  & 0.3783 & 10$\pm$2   & 3$\pm$2     &  36.8  &  46.0& SF \\  	
  574\_C  &   38.61 &  09:50.9  & 0.3928 & 10$\pm$2   & 2$\pm$1     &  76.7  &  87.9& SF \\  	
  575\_C  &   42.63 &  09:51.2  & 0.3953 & 11$\pm$2   & 2$\pm$1     &  80.8  &  94.5& SF \\  	
  582\_C  &   41.37 &  09:51.4  & 0.3957 & 15$\pm$4   & 3$\pm$2     &  37.7  &  44.6& SF \\  	
  612\_C  &   34.45 &  09:52.2  & 0.3945 & 25$\pm$6   & 12$\pm$6    &  53.1  &  75.4& BLAGN \\  	
  620\_C  &   44.70 &  09:52.9  & 0.4062 & 9$\pm$5    & 2$\pm$4     &  9.1  &  11.1& SF \\		
  625\_C  &   41.90 &  09:53.3  & 0.3964 & 5$\pm$2    & 1$\pm$1     &  29.3  &  32.6& SF \\  	
  640\_C  &   28.87 &  09:56.8  & 0.3930 & 11$\pm$3   & 3$\pm$5     &  27.5  &  34.4& SF \\  	
  645\_C  &   28.88 &  09:57.3  & 0.3794 & 20$\pm$5   & 3$\pm$2     &  89.5  &  103.2& SF \\ 	
  658\_C  &   38.39 &  09:59.1  & 0.3950 & 7$\pm$2    & 6$\pm$3     &  35.4  &  64.9& NLAGN \\  	
  659\_C  &   36.85 &  10:00.2  & 0.3962 & 13$\pm$3   & 14$\pm$4    &  60.3  &  118.0& NLAGN \\ 	
  662\_C  &   39.02 &  10:02.3  & 0.4181 & 9$\pm$3    & -      &  30.9  &  - & SF \\  	
  675\_C  &   32.00 &  10:02.6  & 0.3941 & 72$\pm$15  & 22$\pm$8    &  49.4  &  63.3& SF \\  	
  677\_C  &   16.92 &  10:02.8  & 0.3914 & 51$\pm$19  & 41$\pm$38    &  13.3  &  24.0& BLAGN \\ 	
  687\_C  &   22.02 &  10:05.6  & 0.3917 & 13$\pm$4   & 21$\pm$9    &  19.2  &  51.6& NLAGN \\  	
  714\_C  &   24.51 &  10:06.4  & 0.3962 & 27$\pm$6   & 14$\pm$4    &  46.7  &  69.9& SF \\  	
  728\_C  &   31.25 &  10:11.2  & 0.3935 & 15$\pm$4   & 14$\pm$5    &  46.1  &  84.5& BLAGN \\  	
  730\_C  &   28.51 &  10:12.5  & 0.3993 & 7$\pm$3    & 3$\pm$3     &  26.8  &  36.4& SF \\  	
  748\_C  &   31.07 &  10:16.4  & 0.3798 & 24$\pm$5   & 12$\pm$7    &  65.6  &  91.9& SF \\  	
  758\_C  &   24.84 &  10:21.9  & 0.3959 & 22$\pm$6   & 21$\pm$8    &  35.0  &  67.7& NLAGN \\  	
  762\_C  &   32.72 &  10:24.9  & 0.3945 & 10$\pm$3   & 4$\pm$2     &  27.7  &  38.7& SF \\  	
  763\_C  &   25.86 &  10:26.8  & 0.3746 & 15$\pm$8   & 36$\pm$21    &  7.6  &  26.5& NLAGN \\  	
  772\_C  &   25.87 &  10:27.1  & 0.3982 & 16$\pm$5   & 2$\pm$2     &  50.1  &  55.6& SF \\  	
  773\_C  &   33.90 &  10:27.2  & 0.3949 & 30$\pm$7   & 3$\pm$6     &  104.5  &  114.0& SF \\	
  775\_C  &   33.91 &  10:27.7  & 0.3918 & 8$\pm$2    & 1$\pm$1     &  40.7  &  46.9& SF \\  	
  777\_C  &   30.83 &  10:29.1  & 0.3805 & 46$\pm$8   & 14$\pm$4    &  88.6  &  111.1& SF \\ 	
  783\_C  &   30.84 &  10:29.6  & 0.4011 & 6$\pm$3    & 4$\pm$4     &  7.4  &  11.8& SF \\		
  794\_C  &   25.97 &  10:30.1  & 0.3768 & 16$\pm$5   & -	    &  55.9  &  - & SF \\  	
  811\_C  &   37.98 &  10:31.3  & 0.4028 & 4$\pm$1    & 6$\pm$2     &  36.7  &  89.2& NLAGN \\  	
  814\_C  &   28.49 &  10:36.0  & 0.3782 & 7$\pm$3    & -     &  28.8  &  - & SF \\  	
  826\_C  &   37.30 &  10:37.0  & 0.3962 & 33$\pm$7   & 26$\pm$9    &  52.9  &  91.5& NLAGN \\  	
  850\_C  &   12.84 &  10:37.1  & 0.3919 & 64$\pm$13  & 60$\pm$18    &  41.7  &  83.2& NLAGN \\ 	
  869\_C  &   42.74 &  10:43.3  & 0.3948 & 11$\pm$3   & 6$\pm$3     &  30.5  &  48.3& BLAGN \\  	
  872\_C  &   33.63 &  10:45.7  & 0.3935 & 7$\pm$3    & 3$\pm$2     &  12.6  &  19.0& SF \\  	
  875\_C  &   43.49 &  10:47.0  & 0.3899 & 31$\pm$6   & 6$\pm$4     &  60.4  &  71.1& SF \\  	
  882\_C  &   22.10 &  10:49.1  & 0.4052 & 7$\pm$4    & 3$\pm$4     &  8.3  &  12.0& SF \\		
  884\_C  &   28.79 &  10:49.9  & 0.3967 & 14$\pm$3   & 3$\pm$4     &  48.0  &  57.0& SF \\  	
  886\_C  &   28.80 &  10:50.2  & 0.3938 & 7$\pm$2    & 9$\pm$2     &  53.7  &  117.4& NLAGN \\ 	
  888\_C  &   43.31 &  10:51.5  & 0.3888 & 17$\pm$3   & 11$\pm$5    &  52.6  &  90.6& NLAGN \\  	
  889\_C  &   41.50 &  11:02.3  & 0.3952 & 9$\pm$3    & 6$\pm$3     &  24.2  &  40.3& NLAGN \\  	
  907\_C  &   39.14 &  11:02.6  & 0.4190 & 15$\pm$5   & -     &  55.5  &  - & SF \\  	
  916\_C  &   26.71 &  11:02.6  & 0.3836 & 4$\pm$1    & 1$\pm$1     &  26.3  &  33.2& SF \\  	
  923\_C  &   26.71 &  11:02.9  & 0.4035 & 4$\pm$2    & 2$\pm$3     &  7.7  &  11.8& SF \\		
  927\_C  &   37.32 &  11:03.3  & 0.3967 & 34$\pm$7   & 27$\pm$8    &  42.7  &  78.8& NLAGN \\  	
  929\_C  &   26.38 &  11:04.7  & 0.3806 & 32$\pm$11  & 15$\pm$9    &  14.4  &  20.9& SF \\  	
  935\_C  &   26.37 &  11:05.1  & 0.3929 & 9$\pm$3    & 3$\pm$3     &  23.7  &  30.1& SF \\  	
  938\_C  &   36.58 &  11:08.1  & 0.3991 & 4$\pm$1    & 2$\pm$1     &  27.5  &  40.9& SF \\  	
  948\_C  &   35.16 &  11:08.4  & 0.3910 & 14$\pm$3   & 1$\pm$2     &  44.7  &  48.5& SF \\  	
  953\_C  &   35.16 &  11:09.0  & 0.3924 & 5$\pm$2    & 10$\pm$3    &  17.5  &  56.2& NLAGN \\  	
  955\_C  &   43.18 &  11:10.6  & 0.3910 & 9$\pm$3    & 5$\pm$4     &  19.3  &  32.0& NLAGN \\  	
  966\_C  &   22.81 &  11:13.1  & 0.3963 & 43$\pm$9   & 24$\pm$7    &  42.1  &  64.5& SF \\  	
  977\_C  &   30.98 &  11:14.2  & 0.3917 & 8$\pm$2    & 7$\pm$3     &  20.5  &  38.2& NLAGN \\  	
  72\_O   &   25.46 &  11:27.2  & 0.3962 & 27$\pm$4   & 6$\pm$3     &  33.0  &  39.7& BLAGN \\  	
  102\_O  &   24.51 &  11:30.1  & 0.4020 & 28$\pm$3   & 27$\pm$6    &  17.1  &  32.9& BLAGN \\  	
  139\_O  &   27.95 &  11:38.0  & 0.3869 & 14$\pm$2   & 3$\pm$2     &  25.9  &  31.5& SF \\  	
  146\_O  &   33.89 &  11:42.4  & 0.3913 & 20$\pm$2   & 12$\pm$3    &  51.9  &  80.1& BLAGN \\  	
  247\_O  &   25.21 &  11:47.6  & 0.4016 & 24$\pm$11  & 25$\pm$8    &  137.1  &  147.1& SF \\	
  277\_O  &   19.69 &  11:54.1  & 0.3928 & 9$\pm$2    & 6$\pm$3     &  34.6  &  57.0& NLAGN \\  	
  278\_O  &   19.68 &  11:54.4  & 0.3987 & 95$\pm$7   & 113$\pm$8   &  27.6  &  61.1& BLAGN \\  	
  282\_O  &   28.20 &  11:55.5  & 0.3802 & 9$\pm$1    & 1$\pm$1     &  23.7  &  26.4& SF \\  	
  288\_O  &   31.32 &  11:56.7  & 0.3916 & 9$\pm$1    & 6$\pm$1     &  16.8  &  27.2& BLAGN \\  	
  291\_O  &   25.68 &  12:05.4  & 0.3888 & 9$\pm$2    & 6$\pm$3     &  19.3  &  32.4& NLAGN \\  	
  335\_O  &   25.68 &  12:05.7  & 0.3941 & 9$\pm$1    & 8$\pm$3     &  10.6  &  19.9& NLAGN \\  	
  336\_O  &   33.64 &  12:19.7  & 0.3795 & 9$\pm$2    & 2$\pm$2     &  18.0  &  22.8& SF \\  	
  381\_O  &   07.54 &  12:29.9  & 0.3804 & 9$\pm$2    & 1$\pm$2     &  21.2  &  23.9& SF \\  	
  384\_O  &   31.25 &  12:32.1  & 0.3927 & 12$\pm$2   & 2$\pm$1     &  45.5  &  52.7& BLAGN \\  	
  410\_O  &   39.92 &  12:36.7  & 0.3810 & 7$\pm$2    & 6$\pm$3     &  16.8  &  30.9& NLAGN \\  	
  457\_O  &   20.48 &  12:45.4  & 0.3920 & 11$\pm$2   & 8$\pm$2     &  9.9  &  17.2& NLAGN \\		
  485\_O  &   41.88 &  12:45.6  & 0.3935 & 10$\pm$1   & 6$\pm$1     &  31.9  &  50.7& SF \\  	
  501\_O  &   28.37 &  12:46.8  & 0.3790 & 7$\pm$1    & 1$\pm$1     &  11.2  &  12.9& SF \\  	
  567\_O  &   12.54 &  12:54.0  & 0.3970 & 10$\pm$2   & 6$\pm$2     &  25.1  &  39.8& NLAGN \\  	
  568\_O  &   37.59 &  12:54.3  & 0.3791 & 12$\pm$2   & -	    &  28.9  &  - & SF \\  	
  647\_O  &   18.28 &  13:01.0  & 0.3794 & 9$\pm$2    & 1$\pm$2     &  17.5  &  20.6& SF \\  	
  651\_O  &   26.98 &  13:03.7  & 0.3946 & 15$\pm$2   & 9$\pm$3     &  19.2  &  31.7& BLAGN \\  	
  657\_O  &   28.23 &  13:07.8  & 0.3894 & 16$\pm$2   & 7$\pm$3     &  27.3  &  39.5& SF \\  	
  664\_O  &   28.24 &  13:07.9  & 0.3953 & 15$\pm$2   & 9$\pm$2     &  20.8  &  33.7& NLAGN \\  	
  689\_O  &   09.06 &  13:16.1  & 0.4098 & 8$\pm$1    & 1$\pm$2     &  39.6  &  46.3& SF \\  	
  700\_O  &   14.38 &  13:19.9  & 0.3808 & 10$\pm$1   & 10$\pm$3    &  10.8  &  21.6& NLAGN \\  	
  708\_O  &   36.16 &  13:36.4  & 0.3965 & 7$\pm$1    & 2$\pm$2     &  19.7  &  25.5& SF \\  	
  847\_O  &   15.40 &  13:41.7  & 0.3964 & 38$\pm$7   & -	    &  51.9  &  - & SF \\  	
  919\_O  &   27.70 &  13:49.8  & 0.3952 & 14$\pm$2   & 3$\pm$2     &  29.4  &  36.5& SF \\  	
  967\_O  &   23.58 &  13:58.0  & 0.3783 & 8$\pm$1    & -	    &  27.8  &  - & SF \\  	
  998\_O  &   23.62 &  13:58.5  & 0.3916 & 28$\pm$2   & 26$\pm$5    &  38.5  &  71.1& BLAGN \\  	
  1015\_O &   29.19 &  14:02.9  & 0.3937 & 47$\pm$3   & 14$\pm$4    &  40.5  &  51.5& SF \\  	
  1039\_O &   29.14 &  14:03.0  & 0.3961 & 28$\pm$3   & 22$\pm$5    &  23.2  &  41.1& BLAGN \\  	
  1057\_O &   41.25 &  14:04.2  & 0.3934 & 13$\pm$1   & 5$\pm$1     &  32.9  &  44.7& SF \\  	
  1074\_O &   37.75 &  14:12.2  & 0.3711 & 21$\pm$2   & 12$\pm$5    &  33.4  &  51.1& SF \\  	
  1097\_O &   08.54 &  14:58.7  & 0.3947 & 9$\pm$1    & 4$\pm$2     &  53.8  &  76.4& SF \\  	
  1103\_O &   20.14 &  15:10.5  & 0.3960 & 9$\pm$2    & 1$\pm$2     &  29.7  &  34.0& SF \\  	
  1125\_O &   08.95 &  15:10.7  & 0.3914 & 8$\pm$1    & 6$\pm$3     &  28.7  &  48.9& NLAGN \\  	
  1130\_O &   24.16 &  15:27.5  & 0.3919 & 8$\pm$1    & 3$\pm$1     &  34.7  &  46.5& SF \\  	
  1133\_O &   32.95 &  15:29.3  & 0.3717 & 15$\pm$3   & 2$\pm$4     &  23.3  &  25.7& SF \\  	
  1157\_O &   20.26 &  15:32.5  & 0.4042 & 54$\pm$4   & 9$\pm$3     &  69.4  &  80.1& SF \\  	
  1162\_O &   28.36 &  15:36.1  & 0.3848 & 78$\pm$8   & 34$\pm$9    &  29.4  &  41.4& BLAGN \\  	
  1165\_O &   18.65 &  15:42.7  & 0.3950 & 17$\pm$2   & 9$\pm$2     &  15.4  &  23.8& SF \\  	
  1168\_O &   24.75 &  15:48.8  & 0.4005 & 13$\pm$3   & 2$\pm$2     &  19.8  &  22.4& BLAGN \\  	
  1173\_O &   30.79 &  15:53.9  & 0.3901 & 18$\pm$2   & 6$\pm$4     &  38.9  &  50.6& SF \\  	
  1204\_O &   26.10 &  16:47.0  & 0.3948 & 9$\pm$2    & 2$\pm$2     &  32.2  &  38.8& SF \\  	
  1219\_O &   13.33 &  16:54.8  & 0.3820 & 17$\pm$2   & 9$\pm$2     &  22.3  &  33.7& BLAGN \\  	
  1220\_O &   20.28 &  17:03.5  & 0.4006 & 14$\pm$1   & 6$\pm$1     &  35.8  &  50.6& SF \\  	
  		\end{longtable}		
		\end{center}
}

\begin{acknowledgements} 
We acknowledge the support provided by M. Balogh during the preparation of the GLACE proposal to ESO, as well as his useful suggestions to improve the quality of this paper. We also acknowledge the anonymous referee.
We acknowledge financial support from Spanish MINECO under grant AYA2014-29517-C03-01 and AYA2011-29517-C03-02. EJA acknowledges support from MINECO under grant AYA2013-40611-P. JMRE acknowledges support from MINECO under grant AYA2012-39168-C03-01. We acknowledge support from the Faculty of the European Space Astronomy Centre (ESAC). IRS acknowledges support from STFC GT/I001573/1, a European Research Council Advanced Programme Dustygal (321334) and a Royal Society/Wolfson Research Merit Award.
Based on observations made with the Gran Telescopio Canarias (GTC), instaled in the Spanish Observatorio del Roque de los Muchachos of the Instituto de Astrofísica de Canarias, in the island of La Palma.
This publication makes use of data products from the Two Micron All Sky Survey, which is a joint project of the University of Massachusetts and the Infrared Processing and Analysis Center/California Institute of Technology, funded by the National Aeronautics and Space Administration and the National Science Foundation.

\end{acknowledgements}

\end{document}